\documentclass[useAMS,usenatbib]{mn2e}
\pdfoutput=1
\usepackage{graphicx}

\def\aj{AJ}%
\def\actaa{Acta Astron.}%
\def\apj{ApJ}%
\def\apjs{ApJS}%
\def\aap{A\&A}%
\def\aaps{A\&AS}%
\def\mnras{MNRAS}%
\def\na{New A}%
\def\pasp{PASP}%
\def\mnrasa{MNRAS, accepted, arXiv:1407.8091}%

\begin{document}

\title[SMC star clusters]{Bruck\,88 : a young star cluster with an old age resemblance
in the outskirts of the Small Magellanic Cloud}

\author[A.E. Piatti]{Andr\'es E. Piatti$^{1,2}$\thanks{E-mail: 
andres@oac.uncor.edu}\\
$^1$Observatorio Astron\'omico, Universidad Nacional de C\'ordoba, Laprida 854, 5000, 
C\'ordoba, Argentina\\
$^2$Consejo Nacional de Investigaciones Cient\'{\i}ficas y T\'ecnicas, Av. Rivadavia 1917, C1033AAJ,
Buenos Aires, Argentina \\
}

\maketitle

\begin{abstract}
We present spectroscopic and photometric results for the Small Magellanic Cloud (SMC) cluster Bruck\,88.
From the comparison of the cluster integrated spectrum  with 
template cluster spectra we found that the Milky Way 
globular cluster template spectra  are the ones which best resemble it. However, 
the extracted cluster colour magnitude diagram  reveals that 
Bruck\,88 is a young cluster (log($t$) = 8.1 $\pm$ 0.1).
The derived cluster age is compatible with the 
presence of a Bright Red Giant (BRG) star located $\sim$ 2.6 arcsec in the sky from the cluster centre. 
We serendipitously observed HW\,33, a star cluster located $\approx$ 3 arcmin to the south-east from 
Bruck\,88. We obtained for the cluster the same age than Bruck\,88  and surprisingly, a BRG star located 
within the cluster radius also appears to 
be compatible with the cluster age. We estimated the MK type of the BRG star in the Bruck\,88 field to 
be in the range G9\,II/Ib – K1\,III. By combining 
the spectrum of a star within this MK type range  with a 100-150 Myr template cluster 
integrated spectrum, we found that a proportion 85/15 in the sense BRG/template 
results in a spectrum which best resembles that of Bruck\,88. This result confirms that a BRG star dominates
 the cluster integrated spectrum, so that it 
causes the globular cluster appearance of its integrated light.
\end{abstract}

\begin{keywords}
techniques: photometric, spectroscopic -- galaxies: individual: SMC -- Magellanic 
Clouds -- galaxies: star clusters. 
\end{keywords}

\section{Introduction}

The Small Magelllanic Cloud (SMC) has long been known as a galaxy in which a population of
genuine old star clusters does not appear to exist. Up to date, the handful of oldest SMC star
clusters consists in only four objects, namely: NGC\,121 \citep[10.6 Gyr, ][]{detal01}, HW\,42 
\citep[9.3 Gyr, ][]{p11c}; K\,3 \citep[9.0 Gyr, ][]{dh98}; and NGC\,361 \citep[8.1 Gyr, ][]{metal98}.
The remaining small handful of relatively old clusters consists in 12 objects with ages around 
(6 $\pm$ 1) Gyr \citep{petal07,p11c,p12a}.

%From this scenario, some questions arise unavoidable: Is there any genuine old star cluster in the 
%SMC? When did the oldest SMC star clusters form? Have the oldest star clusters been stripped by the 
%Large Magellanic Cloud (LMC)/Milky Way (MW)? When did the first interactions between the SMC/LMC/MW 
%occur?, among others. To answer these questions we should be aware that we have studied in detail 
%the whole SMC cluster population.
%, which is a challenge that requires much more work and effort in 
%order to reach success. 
%Another possibility is to take advantage of theoretical speculations, 
%partially supported by observational results, to infer some conclusions. But it would be even more 
%exciting to identify a single genuine old star cluster in the SMC. If such a discovery were confirmed,
%it would shed light over our present knowledge about the SMC formation and its chemical and dynamical 
%evolutions. 
%However, the discovery of a single old star cluster in the SMC might be very improbable 
%to happen, considering that different campaigns have been carried out until the present and 
%no genuine old cluster has been identified. 
%After the enormous effort invested in this issue, it appears that 
%the general consensus is that the SMC does not harbor genuine old clusters. 

During several years we carried out a spectroscopic campaign aiming at identifying old cluster 
candidates in the L/SMC from their integrated spectra. Our search was based on the comparison of the 
observed integrated continuum energy distribution of the selected targets with the integrated spectra 
library of Galactic globular clusters (GGCs) built by \citet{ba86a}. From this campaign two old globular LMC clusters have been 
identified \citep{detal99}, which have been later confirmed as such \citep{mg04}. Another old 
cluster candidate was identified in the SMC, Bruck\,88, for which we devote here a detailed study. As 
far as we are aware, this cluster has not been previously studied in detail. However, its relative 
position in the outskirts of the galaxy makes it also a potential old cluster candidate (see Fig. 1).
% shows 
%the spatial distribution of clusters catalogued by \cite{betal08} in the SMC, where the four %oldest 
%SMC clusters have been highlighted with open boxes, the relatively old clusters with open 
%triangles, 
%and Bruck\,88 with an open circle. The cluster placed very close to Bruck\,88 to the east-
%south
%is HW\,33. 

The paper is organised as follows: 
The integrated spectroscopic data obtained for Bruck\,88 is presented in Section 2; while the 
cluster fundamental parameters coming from its integrated spectrum are derived in Section 3. Section 4 
and 5 deal with the photometric data and the astrophysical properties estimated for the cluster, 
respectively. In Section 6 we discuss the spectroscopic and photometric results, and summarize the 
conclusions in Section 7.

\section{Spectroscopic data collection and reduction}
 
%The main objective of obtaining an integrated spectrum for Bruck\,88 is to derive estimates of its 
%age and metallicity. 
The compact nature of the cluster - 0.45 arcmin in diameter \citep{betal08} 
- makes it a good target for integrated spectroscopy. The observations were carried out at Cerro 
Tololo Inter-american Observatory (CTIO) through the 2003B-0447 program (PI: A.E. Piatti) with the 
1.5-m telescope during a run in September 18-21, 2003. We employed a CCD detector containing 
a Loral chip of 1200 $\times$ 800 pixels attached to the Cassegrain spectrograph, the size of each 
pixel being 15 $\mu$m $\times$ 15 $\mu$m; one pixel corresponds to 1.3 arcsec on the sky. The slit 
was set in the east-west direction and the observations were performed by scanning the slit across 
the objects in the north-south direction in order to get a proper sampling of cluster stars. The long 
slit corresponding to 7.6 arcmin on the sky, allowed us to sample regions of the background sky. We 
used a grating of 300 lines mm$^{\rm -1}$, producing an average dispersion in the observed region 
of $\approx$ 192 \AA/mm (2.88 \AA/pixel). The spectral coverage was $\sim$ (3500-6800)\AA. A slit 
width of 3.5 arcsec provided a resolution (FWHM) of $\approx$ 11.5 \AA, as deduced from the comparison
lamp lines. Three exposures of 15 minutes were taken for the cluster. The standard stars LTT\,7379,
LTT\, 7987, LTT\,9239, and EG\,21 \citep{sb83} were observed for flux calibrations. Bias, dome and
twilight sky flat fields were taken and employed for the instrumental signature calibrations.

The reduction of the spectra was carried out with the IRAF\footnote{IRAF is distributed by the 
National Optical Astronomy Observatories, which is operated by the Association of Universities for 
Research in Astronomy, Inc., under contract with the National Science Foundation.} package following 
the standard procedures. Summing up, we applied bias and flat field corrections, performed
sky subtraction, extracted and wavelength calibrated the spectra, and 
%we subtracted the bias and used flat-field frames previously 
%combined to correct the frames for high and low spatial frequency variations.
% We also checked the 
%instrumental signature with the acquisition of dark frames. 
%Then, we performed the background sky 
%subtraction using pixel rows from the same frame, after having cleaned the background sky regions from 
%cosmic rays. 
%We controlled that no significant background sky residuals were present on the resulting 
%spectra. 
%The cluster spectra, which were extracted along the slit according to the cluster size and 
%available flux, were then wavelength calibrated by fitting observed He-Ne-Ar comparison lamp %spectra 
%with template spectra. 
The rms errors involved in these calibrations are on average 0.40 \AA. Finally,
extinction correction and flux calibrations 
%derived from the observed standard stars 
were applied to 
the cluster spectra
%. In addition, cosmic rays on them were eliminated 
and the three individual spectra
combined.

\section{Spectroscopic cluster age and metallicity}

The cluster age was derived by comparing the observed spectrum to template spectra 
with well-determined properties, accomplished in various studies 
\citep[e.g.][and references therein]{petal02b} and made available through the CDS/Vizier catalogue
database at http://vizier.u-strasbg.fr/cgi-bin/VizieR?-source=III/219 \citep{setal02}. 
%The integrated spectra library spreads along the whole age range of known clusters. 
We
assigned a higher weight to the overall continuum than to absorption line equivalent widths, 
because the latter have similar equivalent widths for young and old clusters \citep{ba86a,ba86b}. 
%The method consists in achieving the best possible match between the analysed cluster spectrum and
%a template spectrum of known age and metallicity. 
%In this process we selected from the available
%template spectra the one which minimises the flux residuals, calculated as the normalised difference
%(cluster-template)/cluster. 
Note that differences between cluster and template spectra are expected
to be found due to variations in the stellar composition of the cluster, such as the presence of a
relatively bright star with particular spectral features or contamination of a field star close to
the direction towards the cluster.

Since the continuum distribution is also affected by reddening, we firstly adopted a cluster colour 
excess of $E(B-V)$ = (0.03 $\pm$ 0.01) mag taken from the NASA/IPAC Extragalactic 
Database\footnote{http://ned.ipac.caltech.edu/. NED is operated by the Jet Propulsion Laboratory, 
California Institute of Technology, under contract with the National Aeronautics and Space 
Administration.} (NED). 
We computed for Bruck\,88 the value of the semi-major axis $a$ parallel to the SMC main body
that an ellipse would have if it were centred on the SMC centre \citep[RA = 00h 52m 45s, Dec. = 
−72$\degr$ 49$\arcmin$ 43$\arcsec$ (J2000) ]{cetal01}, had a $b/a$ ratio of 1/2 and one 
point of its trajectory coincided with the cluster position. We obtained $a$ = 3.273$\degr$. This
value is larger than the mean semi-major axes of the four oldest SMC clusters ($a$ = 
(2.97$\pm$0.99)$\degr$) which are predominantly placed in the galaxy outer regions.

%To perform reddening corrections, we used the normal reddening law A$_{\lambda}$ =  0.65 A$_V$ 
%(1/$\lambda$-0.35) \citep{s79}, the relation A$_V$ = 3 $E(B-V)$, and the SPEED spectral analysis 
%software \citep{s88}. We have adopted the above reddening law to keep internal consistency with the 
%database of template spectra. 
From the matching procedure we found that the GGC template spectra
G2 ([Fe/H] = -0.4 dex) and G3 ([Fe/H] = -1.0 dex) are the ones which best resemble the dereddened
cluster integrated spectrum. They resulted in lower and upper envelopes of the cluster integrated
spectrum for some spectral regions with metallic bands. Fig. 2 depicts both the observed and 
dereddened Bruck\,88 integrated spectra, as well as the G2 and G3 template spectra. All the spectra 
have been normalised at $\lambda$ = 5870 \AA\, and shifted by an arbitrary constant for comparison 
purposes. 

\citet[hereafter SP04]{sp04} developed a method to estimate cluster ages from visible integrated
spectra. They defined $S_m$ and $S_h$ as the sums of the equivalent widths (EWs) of the metallic lines 
K(CaII), G band (CH) and MgI, and of the Balmer lines H$\delta$, H$\gamma$ and H$\beta$, respectively. 
As they shown, $S_m$ and $S_h$ prove to be useful in the discrimination of old, intermediate-age, and 
young systems. Also, SP04 defined diagnostic diagrams (DDs) involving $S_h$ and $S_m$ with a view to 
discriminating cluster ages for systems younger than 10 Gyr and metallicities for systems older than 
10 Gyr.

We first defined the continuum in the spectrum of Bruck\,88 according to the criteria outlined by 
\citet{ba86a} and then we measured EWs within their selected spectral windows, using IRAF task {\sc splot}.
The boundaries of the spectral windows and their principal absorbers are indicated in \citet{ba86a}. The 
resulting EW measurements (\AA) are shown in Table 1  where the errors come from tracing
slightly different continua. Then, by using the DDs we found that $S_m$ and $S_h$ fall in the region
of old clusters ([Fe/H]$\le$-1.4 dex).  On the other hand, if we compared the individual EWs
listed in Table 1  for Bruck\,88 with those for the G2, G3 and Yf (see Sect. 6) templates, we would
conclude from K(CaII), G band, H$\beta$ and MgI that Bruck\,88 is an old cluster, from H$\delta$ and
H$\gamma$ that it is an intermediate-age cluster, and from H(CaII)+H$\epsilon$ that the cluster
could either be old or young ($\sim$ 100 Myr). The difference between K(caII) and H(CaII)+H$\epsilon$
feaures in old and young objects is a well-known behaviour studied by \citet{r84,r85}.
%As can be seen, from the templates' ages and the EWs analysis
%suggest that Bruck\,88 is an old cluster candidate. 
Nevertheless, the suggested age points to the need of 
a high-quality cluster colour-magnitude diagram 
(CMD), where its Red Giant Branch (RGB), Red Clump (RC), Main Sequence Turnoff (MSTO) and fainter Main 
Sequence (MS) stars can be clearly distinguished.

\section{SDSS \lowercase{$g,i$} photometry}

In order to confirm whether Bruck\,88 belongs to the group of the oldest known clusters in the 
SMC, we built the cluster CMD from $g$ and $i$ images obtained with the Gemini south telescope and the 
GMOS-S instrument (scale = 0.146 arcsec per (2$\times$2 binned) pixel) in the night of January 5th, 
2014. The detector is a 3$\times$1 mosaic of 2K$\times$4K EEV CCDs, yielding a field-of-view (FOV) of 
$\sim$ 5.5$\arcmin$$\times$5.5$\arcmin$. We obtained 4$\times$(30 (short) + 280 (long)) sec exposures 
with the $g$ and $i$ filters, respectively, through program GS-2013B-60 (PI: Piatti). The data were
obtained with an excellent seeing (0.48$\arcsec$ to 0.70$\arcsec$ FWHM), under photometric conditions, 
and with airmass between 1.41 up to 1.51.

The data reduction followed the procedures documented in the Gemini Observatory 
webpage\footnote{http://www.gemini.edu} and utilised the {\sc gemini/gmos} package in IRAF. We 
performed overscan, trimming, bias subtraction, flattened all data images, etc.
%, once the
%calibration frames (zeros and flats) were properly combined. 
Observations of photometric standard 
stars (E5$\_$b and GD$\_$108 standard fields) chosen from the standard star catalog calibrated 
directly in the SDSS system (Smith et al. 2014, http://www-star.fnal.gov) were included in the 
baseline calibrations for GMOS. 
%We used the {\sc apphot} task within IRAF to measure the instrumental
%magnitudes of the standard stars, which were distributed over an area similar to that of the GMOS 
%array, so that we measured magnitudes of standard stars in each of the three chips. 
The relationships 
between instrumental and standard magnitudes, fitted with the {\sc fitparams} IRAF routine, resulted 
to be:

\vspace{0.5cm}
%\begin{equation}
$g = -3.366\pm 0.012 + g_{std} + (0.080\pm 0.010)\times X_g
- (0.011\pm 0.013)\times (g-i)_{std}$ (rms = 0.035) (1)
%\end{equation}

\vspace{0.3cm}

%\begin{equation}
$i = -2.937\pm 0.016 + i_{std} + (0.018\pm 0.010)\times X_i
+ (0.013\pm 0.018)\times (g-i)_{std}$ (rms = 0.038) (2)
%\end{equation}

\vspace{0.5cm}

\noindent where $X$ represents the effective airmass. 

Bruck\,88 photometry was performed using the star-finding and point spread function (PSF) fitting 
routines in the {\sc daophot/allstar} suite of programs \citep{setal90}. We derived a quadratically 
varying PSF per image by fitting $\sim$ 60 least contaminated stars.
%, once the neighbours were eliminated using a 
%preliminary PSF derived from the brightest, least contaminated 20-30 stars. 
%Both groups of PSF stars 
%were interactively selected. 
We then used the {\sc allstar} program to apply the resulting PSF to the
identified stellar objects
% and to create a subtracted image which was used to find and measure 
%magnitudes of additional fainter stars. This procedure was repeated three times for each frame. 
%Finally, we computed aperture corrections from the comparison of PSF and aperture magnitudes by using 
%the neighbour-subtracted PSF star sample. 
%After deriving the photometry for all detected objects in
%each filter, a cut was made on the basis of the parameters returned by {\sc daophot}.
Only objects 
with $\chi$ $<$2, photometric error less than 2$\sigma$ above the mean error at a given magnitude, and
$|$SHARP$|$ $<$ 0.5 were kept in each filter, and then the remaining objects in the $g$ and $i$ lists 
were matched with a tolerance of 1 pixel and raw photometry obtained. 
%We gathered all the independent instrumental magnitudes using the stand-alone {\sc daomatch} and 
%{\sc daomaster} programs\footnote{Program kindly provided by P.B. Stetson.}. 
%We thus produced one 
%data set containing the $x$ and $y$ coordinates for each star, and four different ($g$,$g-i$) pairs; 
%($g$,$i$) shorter and longer exposures being coupled separately. 
The gathered photometric information 
was standardised using equations (1) to (2).
%, and finally the standard magnitudes and colours of stars
%measured several times were averaged. 
%The resulting standardised photometric table consists of  %We 
%adopted the photometric errors provided by {\sc allstar} for stars with only one measure. 
Table 2 
provides this information: a 
running number per star, equatorial coordinates, the averaged $g$ magnitudes and $g-i$ colours, their 
respective rms errors $\sigma(g)$ and $\sigma(g-i)$, and the number of observations per star.
Only a portion of it is shown here for guidance regarding its form and 
content. The whole content of Table 2 is available in the online version of the journal in Oxford 
journals, at http://access.oxfordjournals.org.

\section{Analysis of the Colour-magnitude diagram}

In Fig. 3 we show the CMD of stars measured in the field of Bruck\,88 with errors in $g$ and $g-i$
smaller than 0.1 mag. The behaviour of $\sigma$($g$) and $\sigma$($g-i$) is represented by error bars 
in the right-hand side of the figure. As can be seen, the most obvious traits are the long MS which 
extends over a range of approximately 8 mags in $g$, a populous Sub-Giant branch (SGB), a RC and a RGB. 
The RC is not tilted so that  differential reddening can be assumed to be negligible along the 
line of sight. 

In order to obtain a circular extracted CMD where the fiducial features of the cluster are 
clearly seen, we used the coordinates of the cluster centre and its radius given by \citet{betal08}.
Fig. 4 shows the resulting extracted CMD with all the stars measured within the cluster circle, 
which reveals that Bruck\,88 is a young cluster, although some contamination from field stars is 
unavoidable particularly at the fainter magnitude regime. This result is opposite to that 
found from the cluster integrated spectrum, and makes us to wonder about the origin from which a young
(blue MS) cluster appears much older (redder) to its integrated light.  We discard the possibility of 
field SGB-RC-RGB stars contamination in the cluster integrated spectrum, since the extracted CMD (see 
Fig. 4) was built from stars distributed throughout an area that covers a sky region similar to that 
surveyed from integrated spectroscopy. We rather think that we are dealing with a case of stochastic 
effects caused by the presence of a Bright Red Giant (BRG; $g$ = 16.301$\pm$0.011, $g-i$ = 
1.120$\pm$0.011) star placed somewhere along the cluster line of sight. This BRG is $\approx$ 2.6 
arcsec in the sky from the cluster centre.

In order to derive the cluster age through the matching of theoretical ischrones to its CMD, we 
assumed a distance modulus equal to that of the SMC ($(m-M)_o$ = 18.90$\pm$0.10 
\citep[60.0$^{+3.0}_{-2.5}$ kpc, ][]{getal10}), the present day galaxy metallicity 
\citep[-0.7 dex][]{pg13} - given the apparent youth of the cluster -, and the reddening adopted in 
Sect. 3, $E(B-V)$ = 0.03 mag. 

%For the cluster distance modulus, we adopted the value of the SMC 
%distance modulus $(m-M)_o$ = 18.90$\pm$0.10 \citep[60.0$^{+3.0}_{-2.5}$ kpc, ][]{getal10} because a 
%difference of 0.1 in log($t$) (the difference between two close isochrones used here) implies a 
%difference of $\sim$ 0.40 mag in $g$, whereas the difference in apparent distance modulus - Bruck\,88 
%could be placed in front of or behind the main body of the SM - could be as large as $\Delta(m-M)_o$ 
%$\sim$ 0.15 mag, if values of 60 kpc and 6 kpc \citep{cetal01} are adopted for the mean distance and 
%line of sight depth of the galaxy, respectively. 

We took advantage of theoretical isochrones computed with core overshooting for the SDSS photometric 
system \citep{betal12} to estimate the cluster age.
%, once they were shifted by $E(g-i)$ and by the
%apparent distance modulus, which were calculated using the $E(g-i)$/$E(B-V)$ = 1.621 and 
%$A_{g}$/$E(B-V)$ = 3.738 ratios given by \citet{cetal89}. 
In the matching procedure with a naked eye, 
%we used seven different isochrones, ranging from slightly younger than the derived cluster age to 
%slightly older. Finally, 
we adopted the cluster age as the age of the isochrone which best reproduced 
the cluster's MS (log($t$) = 8.1 $\pm$ 0.1). We found that isochrones bracketing the derived mean age 
by $\Delta$(log($t$/yr)) = $\pm$0.1 reasonably represent the overall age uncertainty. Note that the 
derived cluster age is compatible with the presence of the BRG star mentioned above within the cluster 
stellar population. Fig. 4 illustrates the result of the isochrone matching procedure.

We serendipitously observed HW\,33 within the GMOS-S FOV, a star cluster located $\approx$ 3
arcmin (52.3 pc) to the south-east from Bruck\,88. We followed the same 
procedure of extracting stars around the cluster centre and within the cluster radius and of
building the cluster CMD. We obtained a cluster reddening of $E(B-V)$ = (0.03$\pm$0.01) mag from the
NED. From the matching of theoretical isochrones ([Fe/H] = -0.7 dex) HW\,33 turns out to be of the same
age as Bruck\,88 (log($t$) = 8.1 $\pm$ 0.1, see Fig. 5) and surprisingly, a BRG star ($g$ = 
16.481$\pm$0.004, $g-i$ = 1.058$\pm$0.004) located within the cluster radius, also appears to be
compatible with the cluster age.

\section{Discussion}

In this Sect. we describe a possible explanation aiming at making compatible the results coming from the 
photometric data and our knowledge of the cluster from its integrated spectrum. We propose that the 
observed cluster integrated spectrum is the result of the combination of a template spectrum for a 
$\sim$ 100 Myr old cluster and the spectrum of a bright giant star. Furthermore, we support the 
possibility that the Bruck\,88's integrated spectrum is dominated by the light of a bright giant star 
which mimics the integrated light of an old stellar aggregate. Indeed, when summing to the 
cluster integrated magnitude (computed from all stars in Fig. 4 with $g$ $>$ 17 mag) the magnitude
of the BRG star, Bruck\,88 becomes 1.6 and 4.0 times brighter in $g$ and $i$,
respectively. In order to confirm such a possibility, we first obtained the 
MK type of the aforementioned BRG star, then we selected a spectrum for that MK type from the stellar 
spectra library built by \cite{jetal84}, and finally we combined it with a integrated template spectrum 
for a 100 Myr old cluster taken from \cite{setal02}. 

For the MK type of the BRG star, we interpolated the \citet{shk82}'s MK type vs $(B-V)_o$ and MK type 
vs $M_V$ relations from the computed intrinsic colours $(B-V)_o$ and visual absolute magnitudes $M_V$ of 
the BRG star. 
We computed $M_V$ by combining the SMC distance modulus $(m-M)_o$ = 18.9, the cluster 
reddening $E(B-V)$ = 0.03, the visual to selective absorption ratio $R$ = $A_V$/$E(B-V)$ = 3.1, and the 
relation for $g-V$ given by \citet{jetal06} through the expression:

\begin{equation}
M_V = 0.124 + g - 0.630 \times (B-V)_o - R \times E(B-V) - (m-M)_o,
\end{equation}

\noindent and the intrinsic $(B-V)_o$ colour by combining the relationships between $(B-V)_o$ versus
$(V-I)_o$ of \citet[][see figure \#26]{cetal93} and that between $(V-I)_o$ versus 
$(g-i)_o$ of \citet{jetal06} given by:

\begin{equation}
(g-i)_o   = 1.481 \times (V-I)_o - 0.536.
\end{equation}

\noindent We obtained $M_V$ = (-3.28 $\pm$ 0.20) mag and $(B-V)_o$ = (1.10 $\pm$ 0.02) mag, which 
lead to an MK type of G9/K0\,II/Ib for the BRG star. 

If the RGB star were not located at the SMC distance but somewhere in front of it, then it should be 
mainly moved in the HR diagram along a vertical line from LC\,II towards LC\,III region, because of the 
low reddening along the line of sight ($E(B-V)$ = 0.03). In case it were a LC\,III star,
its $(B-V)_o$ colour would correspond to a K1 MK type star \citep{shk82} and its $M_V$ would be 0.6 mag,
which corresponds to a distance of 10.2-10.5 kpc (see eq. (1)), depending on whether the $E(B-V)$ is
assumed to be 0.00 mag or 0.03 mag. 

Bearing in mind the possible MK types of the BRG star, we selected spectra of stars with MK types in the 
range G9\,II/I-K1\,III, namely: SAO\,55164 (K0\,III), SAO\,77849 (K2\,III), HD\,249384 (G8\,II), 
HD\,250368 (G9\,II), HD\,187299 (G5\,I) and HD\,186293 (K0\,I) from the \citet{jetal84}'s library.
We normalised all spectra at $\lambda$ = 5870 \AA\,  in order to compare them to those from the
cluster integrated spectra library.
%Note that the spectra for the LC\,I class stars are included for 
%comparison purposes since those stars would be located far away behind the SMC, a possibility that we discarded 
%in our analysis. 
We found that all selected spectra are tightly similar to each other with small 
difference in some metalic bands. When individually combining them with the template cluster integrated 
spectrum  Yf \cite[100-150 Myr,][]{setal02}, we found that a proportion 85/15 in the sense BRG/template 
(BRG + template = 100) results in a spectrum which best resembles that of Bruck\,88. Figs. 6 and 7 
illustrate the combination process and the resulting comparison with Bruck\,88's integrated spectrum. 
On the other hand, EWs of some selected spectral features for these spectra combinations (see Table 1) 
seem to be more similar to those of Bruck\,88 than of G2, G3 and Yf, respectively.
This result confirms that a RGB star dominates the cluster integrated spectrum, so that it causes
the globular cluster appearance of its integrated light (see Fig. 2).

%Finally, we would like to mention that \cite{p11b} found, based on the statistics of catalogued and 
%studied clusters, that a total of seven relatively old/old SMC clusters have not yet studied, and even a 
%smaller number is obtained if the cluster spatial distribution is considered. Since then, only 
%ESO\,51-SC09 has been identified as a relatively old SMC cluster \citep{p12b}. Therefore, further work 
%is needed in order to attain having catalogued the entire SMC old cluster population; the present results 
%being witnesses of so challenging goals.

\section{Scope of the integrated spectroscopy}

The misleading integrated spectrum of Bruck\,88 drove us to investigate whether a similar situation might 
exist for other clusters in the literature. Furthermore, we took advantage of such a study to assess the 
level of accuracy and reliability of age estimates derived from the template matching technique applied in 
Sect. 3. We embarked in such an analysis by firstly searching the available literature for studies 
about star clusters based on integrated spectroscopy. Particularly, we focused on those works that made use
of the template integrated spectra library employed here \citep{setal02}. We found a total of 230 clusters
with published ages obtained from this technique. Table 3 lists the entire cluster sample along with the
age estimates (log(age)$_{\rm spec}$) and the respective references. 

From Table 3, we secondly searched the literature for cluster age estimates derived from the analysis of 
CMDs, which provide ages with uncertainties typically of $\sigma$(log(age)) $\sim$ 0.1 
\citep[see e.g.][]{p10,metal14,p14}. We found 198 clusters which fulfilled our request, and their age values and the 
corresponding references have also been included in Table 3 (log(age)$_{\rm cmd}$). As far as we are aware, 
this cluster sample largely supersedes those previously used to set the performance of the integrated 
spectroscopy in estimating cluster ages from the comparison with ages derived from other methods 
\cite[see e.g.][]{asad13}. We note that two objects have been recently classified as cluster remnants, while 
eleven other objects have been found not to be real physical systems from detailed CMD analyses. All of 
them have been labelled with the word 'remnant' or 'non cluster' in Table 3. This flub from the integrated 
spectroscopy side comes from the fact that the technique is not able to size up whether it deals with the 
presence of a genuine star cluster, or with a chance grouping of stars or with the effect of a non-uniform 
distribution of the interstellar material along the line of sight.

The gathered information in Table 3 was then plotted in Fig. 8, which depictes the relationship between 
both different age estimates. We have drawn the identity relation with a black line, and those shifted by 
$\pm$ 0.1, and $\pm$ 0.3 with dark gray and clear gray lines, respectively. Clusters whose age estimates 
differ ($|$log(age)$_{\rm spect}$-log(age)$_{\rm cmd}$$|$) in 0.1, between 0.1 and 0.3, 0.3 and 0.35, and more 
than 0.4 have been plotted with open, filled clear gray, dark gray, and black circles, respectively.
Bruck\,88 has been represented by a bigger open partially filled circle. As can be seen, the relationship 
is far from being reasonably tight but highly scattered. Indeed, Fig. 9 shows the distribution of clusters
in terms of their age differences (absolute values) using the same coloured-gray scale as in Fig. 8. The
age differences larger than 0.4 (black filled squares) represent $\sim$ 55 per cent of the cluster sample. 
This result suggests that the integrated spectroscopy matching technique should be used with caution, since 
it particularly might misguide when applying it in studies of stellar population synthesis in galaxies 
\citep{aetal07,petal08,tetal10,asad13,metal14b}.

Fig. 8 also suggests that the cluster sample is not homogeneously distributed in age; a trend that is 
quantitatively confirmed by Fig. 10. Thus, in order to produce a more realistic picture of the integrated 
spectroscopy performance, we computed the mean, the median, the standard deviation and the mean error of 
the median as a function of the cluster age. The mean error of the median is more representative of the actual 
dispersion of the age differences, since it takes into account the number of clusters used to compute the 
mean and median ages. Similarly, the median is more appropriate to consider in our analysis, because of the
distribution of clusters in the $|$log(age)$_{\rm spect}$-log(age)$_{\rm cmd}$$|$ versus log(age)$_{\rm cmd}$ 
plane for each age interval is far from being a normal distribution. In general, the mean values of the age
difference (absolute value) resulted much larger than the median ones, which would lead to draw less 
favourable conclusions for the integrated spectroscpocy template matching technique, if we used them for 
assessing its accuracy and realibility. For this reason, we show in Fig. 11 the behaviour of the resulting
median values of the age difference and their respective mean errors. As can be seen, a mean discrepancy
of $\sim$ 0.4 in log(age) dominates the integrated spectroscopy age estimates respect to the literature 
values. The differences are reasonably good ($\sim$ 0.1-0.2) 
for a couple of age intervals. However, they come out from a relatively small statistical cluster sample 
and the respective mean errors are considerably large, except in the case of oldest clusters. Moreover, 
Fig. 11 shows that age intervals with 2-3 times more clusters (log(age)$_{\rm cmd}$ $\sim$ 7.8-
8.2) do not accont for age differences smaller than $\sim$ 0.3. Likewise, there are some age intervals
with particularly much larger age differences (log(age)$_{\rm cmd}$ $\sim$ 6.8-7.2, 8.6-8.8, and
9.2-9.6).

These results point to the need of entering some caveats in the use of the integrated spectroscopy matching
technique to estimate cluster ages as it has been applied until the present. On the one hand, the
integrated spectra library is constrained to certain age ranges, which implies that only ages associated
to the available template spectra can be assigned to any integrated spectrum under study. In order to 
enlarge and improve the sample of template spectra, it is required to obtain integrated spectra of clusters
with well-known fundamental parameters. However, several works that have used such a technique (see Table 
3) mentioned that integrated spectra of clusters whose ages were estimated from the integrated spectra 
matching procedure were used to define new template spectra or to improve existing ones. This approach
suffers from internal inconsistency and makes any attempt of enlarging the integrated spectra library to 
fall appart. On the other hand, integrated spectra are often affected at different levels by the 
contamination of field stars. This is an issue that unfortunately the integrated spectroscopy cannot 
handle, particularly in relatively crowded fields or in low-surface brightness objects. Finally, the
presence of evolved or peculiar relatively bright stars within the cluster population can also lead
to a disguised appearence of the cluster integrated light. For instance, we found that Pismis\,7 and
Ruprecht\,107, among others, have red supergiant stars which make the clusters appear older than they are
(see Table 3), similarly as it happens with the integrated spectrum of Bruck\,88. We think that these
aspects or a combination of them can explain the discrepancies shown in Fig. 11. Nevertheless, we
foresee that these kind of constraints will be partially solved with the advent of synthetic
integrated spectra. It is a promising field to exploit that deserve much more development, although
the assumption of a particular cluster composite stellar population to combine different synthetic
spectra is unavoidable.

\section{Conclusions}

During several years we carried out a spectroscopic campaign aiming at identifying old cluster candidates 
in the L/SMC from their integrated spectra. 
From that campaign we identified
% NGC\,1928 and 1939 have been
%identified as new old globular LMC clusters. Another old cluster candidate was identified in the SMC, 
Bruck\,88, for which we present the following spectroscopic and photometric results:

%i) We obtained the cluster integrated spectrum from observations performed at CTIO 1.5-m telescope with
%the Cassegrain spectrograph; the spectral coverage being $\sim$ (3500-6800) \AA\, and the resolution (FWHM) 
%$\approx$ 11.5 \AA, respectively. 
i) From the comparison of the cluster integrated spectrum with template 
cluster spectra, 
%with well-determined properties spread along the whole age range of known clusters, 
we found 
that the GGC template spectra G2 and G3 are the ones which best resemble the 
%dereddened 
cluster 
integrated spectrum. 

%ii) We computed for Bruck\,88 the value of the semi-major axis ($a$) that an ellipse would have if it 
%were centred on the SMC centre, had a $b/a$ ratio of 1/2 and one point of its trajectory coincided with 
%the cluster position. We derived $a$ = 3.273$\degr$, which implies that the cluster is placed - within 
%this elliptical framework - in the galaxy outer region. This $a$ value is, in turn,  larger than the 
%mean $a$ values of the four oldest SMC clusters ($a$ = (2.97$\pm$0.99)$\degr$). The relatively low colour 
%excess derived from the NED for the cluster ($E(B-V)$) is also in agreement with its possition in the 
%outskirts of the galaxy.

%iii) In order to confirm whether Bruck\,88 belongs to the group of the oldest known clusters in the SMC, 
%we built the cluster CMD from $g$ and $i$ images obtained with the Gemini south telescope and the  
%GMOS-S instrument. The high-quality photometric data unveiled the existente of a stellar composite 
%population of a long MS which extends over a range of approximately 8 mags in $g$, a populous SGB, a 
%RC and a RGB. The RC is not tilted so that differential reddening can be assumed to be negligible along 
%the line of sight. 

ii) However, the extracted cluster CMD where its fiducial features are clearly seen reveals that 
Bruck\,88 is a young cluster (log($t$) = 8.1 $\pm$ 0.1). 
%According to the matching of the theoretical isochrone which best 
%reproduced the cluster's MS, we adopted an age of log($t$) = 8.1 $\pm$ 0.1, whose error reasonably 
%represents the overall age uncertainty. 
Note that the derived cluster age is compatible with the presence 
of a BRG star located $\sim$ 2.6 arcsec away from the direction towards the cluster centre. This result is 
opposite to that found from the cluster integrated spectrum, and makes us to wonder about the origin from 
which a young cluster appears much older to its integrated light.  

iii) We serendipitously observed HW\,33 within the GMOS-S FOV, a star cluster located $\approx$ 3
arcmin (52.3 pc) to the south-east from Bruck\,88. We obtained for the
cluster the same $E(B-V)$  colour excess and age than Bruck\,88  and surprisingly, a BRG star located
within the cluster radius also appears to be compatible with the cluster age.

%iv) We propose the possibility that the Bruck\,88's integrated spectrum is dominated by the light of a 
%bright giant star which mimics the integrated light of an old stellar aggregate. To support such a possibility, 
iv) We 
estimated the MK type of the Bruck\,88's BRG star to be in the range G9\,II/Ib - K1\,III.
%, depending on whether
%the reddening is taken into account and the star is assumed to be located at the SMC distance
%or somewhere in front of it. 
By combining the spectrum of a star within this MK type range  
with a 100-150 Myr template cluster integrated spectrum, we found that a proportion 85/15 in the sense 
BRG/template (BRG + template = 100) results in a spectrum which best resembles that of Bruck\,88, 
independently of the BRG star LC. This result confirms that a BRG star dominates the cluster integrated 
spectrum, so that it causes the globular cluster appearance of its integrated light.

\section*{Acknowledgements}
Based on observations obtained at the Cerro Tololo Inter-American Observatory (CTIO, Program: 2003B-0447), 
which is operated by the Association of Universities for Research in Astronomy (AURA), Inc., under 
cooperative agreement with the NSF and at the Gemini Observatory (Program: GS-2013B-Q-60), which is 
operated by AURA, Inc., under a cooperative agreement with the NSF on behalf of the Gemini partnership: 
the National Science Foundation (United States), the Science and Technology Facilities Council (United 
Kingdom), the National Research Council (Canada), CONICYT (Chile), the Australian Research Council 
(Australia), Minist\'erio da Ci\`encia, Tecnologia e Inova\c{c}\~ao (Brazil) and Ministerio de Ciencia, 
Tecnolog\'{\i}a e Innovaci\'on Productiva (Argentina). 
I thank the astronomers at CTIO who carried out
the observations. This research has made use of the SIMBAD database, operated at CDS, Strasbourg, France, 
and draws upon data as distributed by the NOAO Science Archive. NOAO is operated by the Association of 
Universities for Research in Astronomy (AURA), Inc. under a cooperative agreement with the National 
Science Foundation.  This work was partially supported by the Argentinian institutions
CONICET and Agencia Nacional de Promoci\'on Cient\'{\i}fica y Tecnol\'ogica (ANPCyT). 
I am grateful for the comments and suggestions raised by the anonymous
referee which helped me to improve the manuscript.

%to be commented before sending to editor
\bibliographystyle{mn2e_new} %style mn.bst
%\bibliography{paper} % your references file.bib
% 
%to be uncommented before sending to editor

%

\clearpage

\begin{table}
\tiny
\caption{EWs (\AA) of selected espectral features.}
\begin{tabular}{@{}lccccccc}\hline
Spectrum   & K(CaII) &  H(CaII)+H$\epsilon$ & H$\delta$ & G band (CH) & H$\gamma$ & H$\beta$ & MgI \\\hline
Bruck\,88  & 7.1$\pm$0.5 &  8.7$\pm$0.4 & 3.8$\pm$0.5 & 3.3$\pm$0.3 & 5.2$\pm$0.4 & 3.8$\pm$0.3 & 2.6$\pm$0.3 \\
G2         & 8.8$\pm$0.3 &  7.7$\pm$0.4 & 1.3$\pm$0.1 & 4.4$\pm$0.2 & 1.3$\pm$0.2 & 3.3$\pm$0.2 & 2.7$\pm$0.1 \\
G3         & 6.5$\pm$0.3 &  6.7$\pm$0.3 & 1.8$\pm$0.1 & 3.6$\pm$0.1 & 1.8$\pm$0.1 & 3.4$\pm$0.2 & 1.6$\pm$0.1 \\
Yf       & 1.9$\pm$0.2 &  7.3$\pm$0.2 & 7.9$\pm$0.2 & 0.8$\pm$0.2 & 8.3$\pm$0.1 & 9.2$\pm$0.3 & 0.5$\pm$0.1 \\
0.85$\times$Yf+0.15$\times$G9II & 7.8$\pm$0.2 &  8.7$\pm$0.3 & 2.2$\pm$0.1 & 4.6$\pm$0.1 & 4.4$\pm$0.4 & 3.3$\pm$0.3 & 3.0$\pm$0.2 \\
0.85$\times$Yf+0.15$\times$K1III &7.1$\pm$0.2 &  8.0$\pm$0.3 & 3.2$\pm$0.3 & 6.5$\pm$0.3 & 3.0$\pm$0.3 & 3.5$\pm$0.2 & 3.2$\pm$0.2 \\
\hline
\end{tabular}
\end{table}

\begin{table}
\tiny
\caption{CCD $gi$ data of stars in the field of Bruck\,88.}
\begin{tabular}{@{}lccccccc}\hline
Star & RA(J2000)  & DEC(J2000) & $g$ & $\sigma$($g$) & $g-i$ & $\sigma$$(g-i)$ & n \\
     & (h:m:s) & ($\deg$ $\arcmin$ $\arcsec$) & (mag) & (mag) & (mag) & (mag)  \\\hline
-    &   -     &   -     &  -    &  -    &   -   &   -     \\
     20 & 00:57:18.373 & -70:49:07.82 &  22.029 &   0.013 &   2.216   & 0.014 &  8\\
     21 & 00:57:18.828 & -70:49:07.36 &  19.943 &   0.005 &   0.368   & 0.007 &  8\\
     22 & 00:56:48.942 & -70:49:07.23 &  21.428 &   0.006 &   0.684   & 0.012 &  8\\
-    &   -     &   -     &  -    &  -    &   -   &   -     \\
\hline
\end{tabular}
\end{table}

\begin{table*}
\begin{minipage}{126mm}
\caption{Age estimates derived from CMDs and integrated spectra.}
\begin{tabular}{@{}lcccclcccc}\hline
Cluster & log(age)$_{\rm spec}$  & Ref. & log(age)$_{\rm cmd}$ & Ref. &
Cluster & log(age)$_{\rm spec}$  & Ref. & log(age)$_{\rm cmd}$ & Ref.\\\hline
Alessi\,14      & 8.7    & 10    &non cluster  & 11 &  NGC\,1856       & 8.45   &  1    & 8.45  &  2\\
Alessi\,15      & 9.0    & 10    & 9.25  & 11 &  NGC\,1859       & 8.70   &  9    & 8.1   &  3\\
Alessi\,16      & 9.50   & 10    & 9.05  & 11 &  NGC\,1863       & 6.90   &  1    & 7.7   &  4\\
Basel\,18       & 7.7    & 31    & 7.6   & 11 &  NGC\,1870       & 7.7    & 59    & 7.5   & 58\\  
Berkeley\,75    & 9.3    & 38    & 9.5   & 50 &  NGC\,1887       & 7.8    & 22    & 8.1   &  3\\
Berkeley\,77    & 9.55   & 30    & 8.8   & 15 &  NGC\,1894       & 8.1    & 59    & 7.75  & 58\\
Berkeley\,80    & 8.8    & 31    & 8.6   & 51 &  NGC\,1897       & 8.65   & 43    & --   &  --\\
BH\,55          & 8.8    & 14    & 9.05  & 11 &  NGC\,1902       & 7.8    & 59    & 8.0   &  3\\
BH\,58          & 8.6    & 14    & --   &  -- & NGC\,1903       & 7.8    &  1    & --   &  --\\
BH\,72          & 8.7    & 10    & 9.1   & 11 &  NGC\,1905       & 8.5    & 43    & --   &  --\\
BH\,80          & 6.65   & 38    & 6.7   & 11 &  NGC\,1913       & 7.8    & 59    & 7.50  & 58\\
BH\,87          & 8.0    & 38    & 8.4   & 11 &  NGC\,1920       & 6.7    &  9    & --   &  --\\
BH\,90          & 8.55   & 14    & 7.95  & 11 &  NGC\,1932       & 8.6    & 59    & 8.1   &  3\\
BH\,92          & 8.55   & 14    & 7.6   & 11 &  NGC\,1940       & 7.8    & 59    & 8.0   &  3\\
BH\,121         & 6.6    & 14    & 6.65  & 53 &  NGC\,1943       & 8.45   & 63    & 7.50  & 58\\
BH\,132         & 8.0    & 38    & --   &  -- &  NGC\,1944       & 7.8    & 22    & --   &  --\\
BH\,151         & 6.45   & 31    & --   &  -- &  NGC\,1971       & 7.7    & 59    & 7.8   & 58 \\
BH\,202         & 9.0    & 24    & 8.05  & 11 &  NGC\,1972       & 7.85   & 22    & 7.6   & 58\\
BH\,205         & 7.0    & 14    & 7.1   & 11 &  NGC\,1983       & 6.65   &  1    & 7.45  & 44\\
BH\,217         & 7.55   & 14    & 7.65  & 11 &  NGC\,1984       & 6.65   &  1    & 7.8   &  3\\
Bochum\,2       & 6.7    & 31    & 6.7   & 52 &  NGC\,1986       & 7.8    & 22    & --   &  --\\
Bochum\,12      & 7.5    & 38    & 7.6   & 11 &  NGC\,1994       & 6.50   &  1    & 7.7   &  3\\
Bochum\,14      & 6.45   & 34    & 7.0   & 11 &  NGC\,2000       & 7.6    & 22    & 8.0   &  3\\
Bruck\,50       & 6.6    & 26    & 7.0   & 27 &  NGC\,2002       & 6.65   &  1    & 7.1   &  3\\
Collinder\,258  & 8.0    & 14    & 8.05  & 53 &  NGC\,2011       & 6.65   &  1    & 7.4   &  3\\
Dolidze\,34     & 8.8    & 31    & --   &  -- &  NGC\,2031       & 7.80   &  1    & 8.2   &  5\\
ESO\,065-SC07   & 9.4    & 14    & 9.1   & 11 &  NGC\,2038       & 7.8    & 59    & --   &  --\\
ESO\,211-SC09   & 9.55   & 10    & 9.05  & 11 &  NGC\,2053       & 7.85   & 22    & 8.25  &  3\\
ESO\,260-SC06   & 9.50   & 10    & 8.80  & 11 &  NGC\,2065       & 7.8    &  1    & --   &  --\\
ESO\,277-SC04   & 9.0    & 10    & --   &  -- &  NGC\,2095       & 7.6    &  9    & 7.9   &  3\\
ESO\,313-SC03   & 9.50   & 10    & 8.85  & 11 &  NGC\,2097       & 8.9    & 23    & --   &  --\\
ESO\,315-SC14   & 8.70   & 10    & 7.85  & 11 &  NGC\,2118       & 7.8    & 59    & 7.2   &  3\\
ESO\,324-SC15   & 9.00   & 30    &remnant  & 16 &  NGC\,2130       & 7.8    & 59    & 7.9   &  3\\
ESO\,332-SC11   & 6.85   & 10    & --   &  -- &  NGC\,2135       & 7.7    & 59    & 7.5   & 58\\
ESO\,371-SC25   & 7.00   & 10    & 9.55  & 11 &  NGC\,2136       & 7.8    & 23    & 7.3   &  3\\
ESO\,429-SC02   & 6.85   & 31    & 8.6   & 11 &  NGC\,2137       & 9.0    & 22    & 8.1   &  3\\
ESO\,429-SC13   & 8.0    & 14    & --   &  -- &  NGC\,2140       & 7.8    & 22    & 8.0   &  3\\
ESO\,445-SC74   & 9.45   & 31    & 8.2   & 11 &  NGC\,2155       & 9.00   &  1    & 9.55  &  6\\
ESO\,492-SC02   & 6.90   & 14    &non cluster  & 11 &  NGC\,2156       & 6.90   &  1    & 7.9   &  3\\
ESO\,502-SC19   & 9.00   & 10    &remnant  & 13 &  NGC\,2157       & 7.80   &  1    & 7.2   &  3\\
Haffner\,7      &  8     & 38    & 9.2   & 40 &  NGC\,2164       & 6.9    &  1    & 7.3   &  3\\
Hodge\,9        & 8.9    & 23    & --   &  -- &  NGC\,2166       & 8.5    & 43    & --   &  --\\
Hogg\,3         & 7.9    & 38    &non cluster  & 39 &  NGC\,2172       & 6.90   & 30    & 7.9   &  3\\
Hogg\,9         & 8.45   & 31    & 7.45  & 11 &  NGC\,2173       & 6.90   &  1    & 9.3   &  7\\
Hogg\,10        & 7.5    & 31    & 7.0   & 33 &  NGC\,2181       & 8.6    & 43    & --   &  --\\
Hogg\,11        & 8.5    & 38    & 7.1   & 11 &  NGC\,2197       & 8.6    & 43    & --   &  -- \\
Hogg\,12        & 7.9    & 38    & 7.5   & 42 &  NGC\,2213       & 8.70   &  1    & 9.25  &  7\\
Hogg\,14        & 8.45   & 24    & 8.1   & 11 & NGC\,2249       & 8.45   &  1    & 9.15  &  7\\
Hogg\,15        & 6.65   & 14    & 7.3   & 18 &  NGC\,2311       & 8.45   & 31    & 8.0   & 32\\
Hogg\,22        & 6.65   & 31    & 6.8   & 11 &  NGC\,2368       & 7.7    & 38    &non cluster  & 39\\
HS\,109         & 7.65   & 22    & 8.0   &  3 & NGC\,2409       & 7.7    & 31    & --   &  --\\
HW\,8           & 7.7    & 61    & 8.0   &  3 &  NGC\,2459       & 9.0    & 10    &non cluster  & 11\\
HW\,73          & 7.7    & 35    & 8.1   & 69 &  NGC\,2587       & 9.0    & 14    & 8.7   & 56\\
HW\,85          & 7.2    & 26    & 9.35  & 29 &  NGC\,2635       & 9.2    & 38    & 8.8   & 11\\
IC\,1611        & 8.1    & 60    & 8.2   &  3 &  NGC\,3255       & 8.45   & 24    & 8.3   & 11\\
IC\,1624        & 7.75   & 35    & 8.3   & 27 &  NGC\,3590       & 7.6    & 24    & 7.5   & 42\\
IC\,1626        & 8.45   & 60    & 8.35  &  3 &  NGC\,4439       & 7.6    & 24    & 7.9   & 11\\
IC\,1641        & 8.5    & 61    & 8.3   &  3 &  NGC\,4463       & 7.5    & 24    & 7.5   & 11\\
Kron\,3         & 9.85   & 61    & 9.95  & 65 &  NGC\,4609       & 8.1    & 24    & 7.7   & 25\\
Kron\,5         & 8.9    & 61    & 9.3   & 64 &  NGC\,5168       & 8.1    & 24    & 8.0   & 11\\
Kron\,6         & 9.3    & 61    & 9.1   & 62 &  NGC\,5281       & 7.5    & 31    & 7.15  & 11\\
Kron\,7         & 9.6    & 61    & 9.55  & 65 &  NGC\,5606       & 6.65   & 38    & 7.1   & 11\\
Kron\,28        & 9.0    & 61    & 9.3   &  6 &  NGC\,6204       & 7.8    & 31    & 7.9   & 11\\
Kron\,34        & 8.45   & 26    & 8.6   & 27 &  NGC\,6268       & 7.8    & 14    & 7.6   & 11\\
Kron\,42        & 7.65   & 61    & 7.8   &  3 &  NGC\,6604       & 6.5    & 31    & 6.8   & 11\\\hline
\end{tabular}
\end{minipage}
\end{table*}

\setcounter{table}{2}
\begin{table*}
\begin{minipage}{126mm}
\caption{continued.}
\begin{tabular}{@{}lcccclcccc}\hline
Cluster & log(age)$_{\rm spec}$  & Ref. & log(age)$_{\rm cmd}$ & Ref. &
Cluster & log(age)$_{\rm spec}$  & Ref. & log(age)$_{\rm cmd}$ & Ref.\\\hline
Lindsay\,5      & 8.9    & 61    & 9.6   & 64 &  Pismis\,7       & 9.65   & 31    & 8.7   & 11\\
Lindsay\,28     & 9.0    & 35    & 9.0   & 29 &  Pismis\,17      & 9.5    & 38    & 7.0   & 11\\
Lindsay\,39     & 7.2    & 61    & 8.0   &  3 &  Pismis\,20      & 6.7    & 31    & 6.9   & 11\\
Lindsay\,51     & 7.2    & 61    & 7.8   &  3 &  Pismis\,21      & 7.90   & 14    &non cluster  & 19\\
Lindsay\,56     & 6.7    & 35    & 7.4   & 27 &  Pismis\,23      & 8.5    & 14    & 8.5   & 11\\
Lindsay\,66     & 7.2    & 61    & 7.4   &  3 &  Pismis\,24      & 6.7    & 31    & 7.0   & 11\\
Lindsay\,95     & 7.8    & 26    & 8.4   & 28 &  Ruprecht\,2     & 9.55   & 31    & --   &  --\\
Lindsay\,114    & 9.75   & 35    & 8.15  & 36 &  Ruprecht\,14    & 9.6    & 10    & --   &  --\\
Lyng\aa\,1      & 8.0    & 31    & 8.0   & 11 &  Ruprecht\,17    & 8.7    & 10    & --   &  --\\
Lyng\aa\,4      & 9.0    & 10    & 9.1   & 11 &  Ruprecht\,38    & 9.0    & 10    & --   &  --\\
Lyng\aa\,11     & 8.65   & 14    & 8.8   & 54 &  Ruprecht\,107   & 9.55   & 24    & 7.5   & 11\\
Markarian\,38   & 7.0    & 31    & 6.75  & 55 &  Ruprecht\,144   & 8.00   & 38    & 8.65  & 20 \\
Melotte\,105    & 8.0    & 70    & 8.55  & 19 &  Ruprecht\,150   & 9.00   & 10    &non cluster  & 12\\
NGC\,121        & 10.1   & 35    & 10.0  & 46 &  Ruprecht\,158   & 8.85   & 14    &non cluster  & 21\\
NGC\,241        & 7.45   & 35    & 8.35  & 37 &  Ruprecht\,159   & 9.3    & 14    & --   &  --\\
NGC\,242        & 7.2    & 35    & 7.8   & 37 &  Ruprecht\,164   & 8.9    & 14    & 9.0   & 11\\
NGC\,256        & 7.4    & 35    & 7.8   &  3 &  Sher\,1         & 7.5    & 24    & 6.7   & 11\\
NGC\,265        & 7.75   & 35    & 8.35  &  3 &  SL\,14          & 7.2    & 22    & 8.3   &  3\\
NGC\,269        & 8.8    & 61    & 8.5   & 27 &  SL\,56          & 8.1    & 23    & 7.5   &  3\\
NGC\,290        & 7.4    & 35    & 7.6   &  3 &  SL\,58          & 7.8    & 22    & 8.0   &  3\\
NGC\,294        & 8.5    & 61    & 8.45  &  3 &  SL\,76          & 7.7    & 22    & 8.1   &  3\\
NGC\,306        & 6.85   & 35    & 7.7   &  3 &SL\,79          & 8.0    & 22    & 8.3   &  3\\
NGC\,330        & 7.6    & 45    & 7.4   &  3 &  SL\,106         & 8.15   &  8    & 7.7   &  3\\
NGC\,411        & 9.0    & 61    & 9.2   & 65 &SL\,116         & 7.7    & 23    & 7.8   &  3 \\
NGC\,416        & 9.75   & 35    & 9.85  & 41 &  SL\,152         & 8.1    & 23    & 8.6   &  3 \\
NGC\,419        & 8.9    & 61    & 9.1   & 65 &  SL\,168         & 7.65   & 22    & 8.3   &  3\\
NGC\,422        & 8.5    & 61    & 8.15  &  3 &  SL\,234         & 6.6    & 22    & 7.95  &  3\\
NGC\,458        & 7.7    & 61    & 8.2   &  6 &  SL\,237         & 7.8    & 45    & 7.3   &  3\\
NGC\,643        & 9.0    & 35    & 9.05  & 29 &  SL\,242         & 7.55   & 60    & 7.9   &  3\\
NGC\,796        & 6.7    & 35    & 8.05  & 36 &  SL\,255         & 7.8    & 22    & 7.8   &  3\\
NGC\,1611       & 8.1    & 26    &non cluster  & 49 &  SL\,256         & 7.8    &  8    & 7.8   &  3\\
NGC\,1626       & 8.45   & 26    &non cluster  & 11 &  SL\,360         & 6.65   & 22    & --   &  --\\
NGC\,1651       & 8.45   &  1    & 9.3   &  2 &  SL\,364         & 7.6    & 22    & 8.0   &  3\\
NGC\,1695       & 8.2    &  9    & 8.0   &  3 &  SL\,386         & 7.8    & 22    & 8.1   &  3\\
NGC\,1696       & 8.65   & 43    & --   &  -- &  SL\,410         & 8.45   &  8    & 8.05  &  3\\
NGC\,1698       & 7.8    &  9    & 8.0   &  3 &  SL\,425         & 9.00   &  8    & 8.2   &  3\\
NGC\,1702       & 8.1    &  9    & 7.8   &  3 &  SL\,463         & 7.7    & 22    & 7.7   &  3\\
NGC\,1704       & 8.45   &  9    & 7.5   &  3 &  SL\,477         & 7.7    & 22    & 7.9   &  3\\
NGC\,1711       & 6.50   &  1    & 7.2   &  3 &  SL\,508         & 7.8    & 59    & --   &  --\\
NGC\,1732       & 7.8    & 22    & 7.7   &  3 &  SL\,543         & 8.45   &  8    & 7.9   &  3\\
NGC\,1754       & 10.1   & 45    & 10.2  & 47 &  SL\,551         & 7.2    & 22    & 7.9   &  3\\
NGC\,1775       & 7.8    & 23    & 7.85  & 58 &  SL\,566         & 7.8    & 22    & 7.85  & 58\\
NGC\,1793       & 7.8    &  9    & 8.0   &  3 &  SL\,624         & 8.7    &  8    & --   &  --\\
NGC\,1804       & 7.7    & 59    & 7.8   &  3 &  SL\,709         & 8.1    & 59    & 7.1   &  3\\
NGC\,1815       & 8.1    &  9    & 7.8   &  3 &  SL\,763         & 7.8    & 22    & 8.0   &  3\\
NGC\,1822       & 8.1    & 22    & 8.0   &  3 &  Trumpler\,15    & 6.5    & 34    & 6.9   & 11\\
NGC\,1828       & 7.7    & 23    & --   &  -- &  Trumpler\,21    & 7.5    & 31    & 7.7   & 11\\
NGC\,1836       & 8.45   & 45    & 8.6   & 48 &  Trumpler\,27    & 6.5    & 14    &non cluster  & 57\\
NGC\,1839       & 8.4    & 45    & 8.1   &  4 &  WG\,1           & 6.85   & 26    & --   &  --\\
NGC\,1850       & 6.90   &  1    & 7.3   &  3 &                  &        &       &       &    \\
\hline
\end{tabular}
\medskip
Ref. (1) \cite{asad13}, (2) \cite{ketal07}, (3) \cite{getal10}, (4) \cite{petal03}, (5) \cite{metal06},
 (6) \cite{petal02a},
 (7) \cite{petal14}, (8) \cite{oetal12}, (9) \cite{metal12}, (10) \cite{betal10}, (11) \cite{detal02}, 
(12) \cite{vetal08}, 
(13) \cite{betal01},
(14) \cite{aetal09}, (15) \cite{letal10}, (16) \cite{petal11}, (17) \cite{aetal00}, (18) \cite{petal02c},
 (19) \cite{pc01},
 (20) \cite{cetal09}, (21) \cite{vetal10}, (22) \cite{tetal09}, (23) \cite{petal08}, (24) \cite{aetal08}, 
(25) \cite{ketal10}, 
(26) \cite{tetal10}, (27) \cite{chetal06}, (28) Piatti et al. (in preparation), (29) \cite{pietal11}, 
(30) \cite{petal08b}, 
(31) \cite{aetal07}, (32) \cite{petal10}, (33) \cite{detal11}, (34) \cite{aetal06}, (35) \cite{aetal02}, 
(36) \cite{petal07b},
 (37) \cite{metal14},
(38) \cite{aetal01}, (39) \cite{pietal11b}, (40) \cite{cetal13}, (41) \cite{setal06},
 (42) \cite{petal10b}, (43) \cite{metal14b},
 (44) \cite{getal10b}, 
\end{minipage}
\end{table*}

\setcounter{table}{2}
\begin{table*}
\begin{minipage}{126mm}
\caption{continued.}
(45) \cite{aetal11}, (46) \cite{detal01}, 
(47) \cite{oetal98}, (48) \cite{petal03b}, (49) \cite{jetal09}, (50) \cite{cetal05}, (51) \cite{tetal10b}, 
(52) \cite{metal06b}, 
(53) \cite{ketal05}, (54) \cite{petal06}, (55) \cite{ssetal12}, 
(56) \cite{petal07c}, (57) \cite{petal12}, (58) \cite{detal02b}, (59) \cite{cetal07}, (60) \cite{tetal07},
 (61) \cite{petal05}, 
(62) \cite{metal02}, (63) \cite{setal06b}, (64) \cite{petal05b}, (65) \cite{dh98}, (66) this paper.
\end{minipage}
\end{table*}

\clearpage

\begin{figure}
%\centerline{\psfig{figure=fig1.ps,width=84mm}}
\includegraphics[width=84mm]{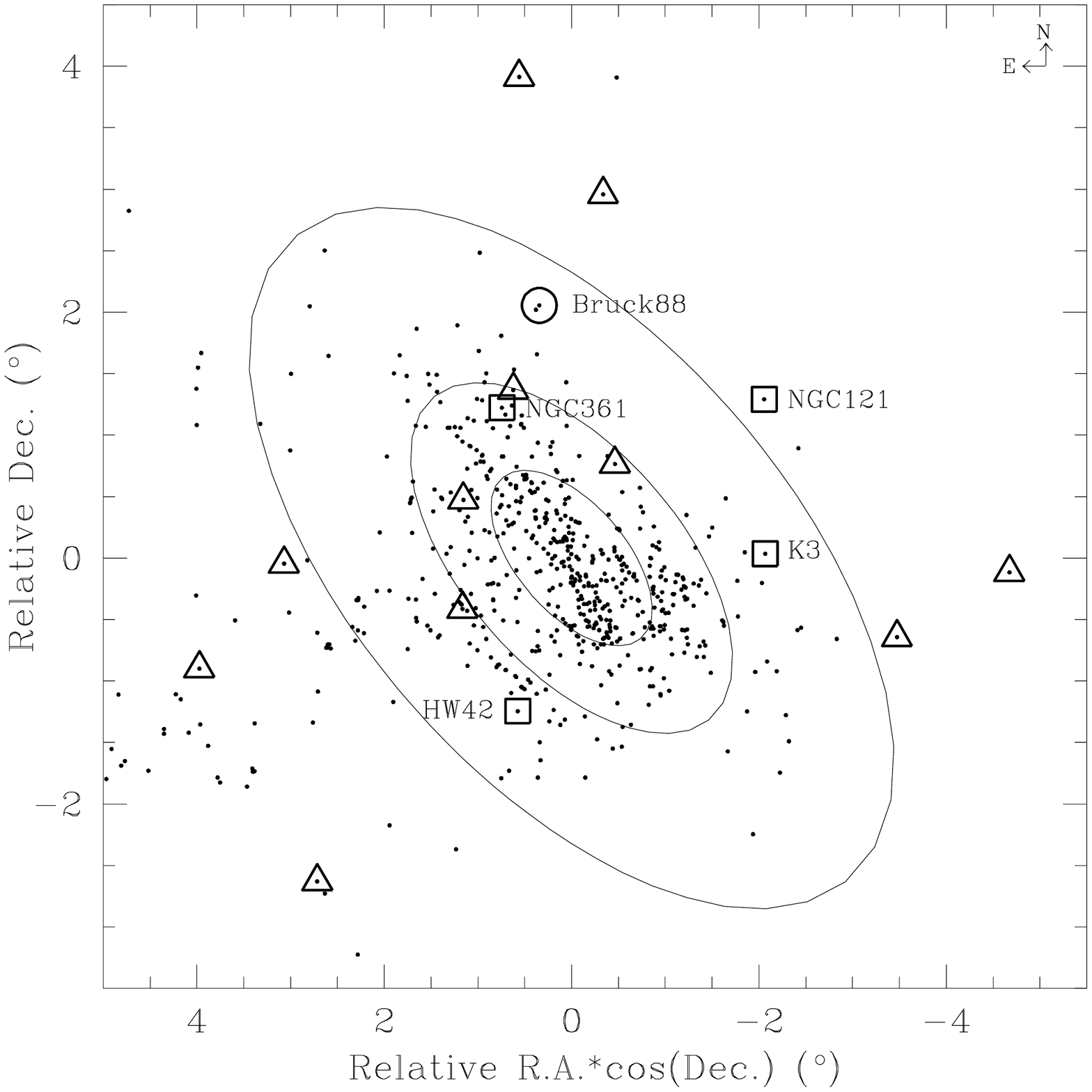}
\caption{Spatial distribution of clusters catalogued by \citet{betal08} in the SMC, where the four oldest 
SMC clusters have been highlighted with open boxes, the relatively old clusters with open triangles, 
and Bruck\,88 with an open circle. The cluster placed very close to Bruck\,88 to the east-south
is HW\,33. 
%Spatial distribution of SMC clusters (dots). The open boxes, triangles and circle
%highlight the position of SMC old, relatively old, and Bruck\,88 clusters, respectively.
Ellipses with semi-major axis of 1$\degr$, 2$\degr$ and 4$\degr$ are overplotted.
}
\label{fig1}
\end{figure}

\begin{figure}
%\centerline{\psfig{figure=fig2.ps,width=84mm}}
\includegraphics[width=84mm]{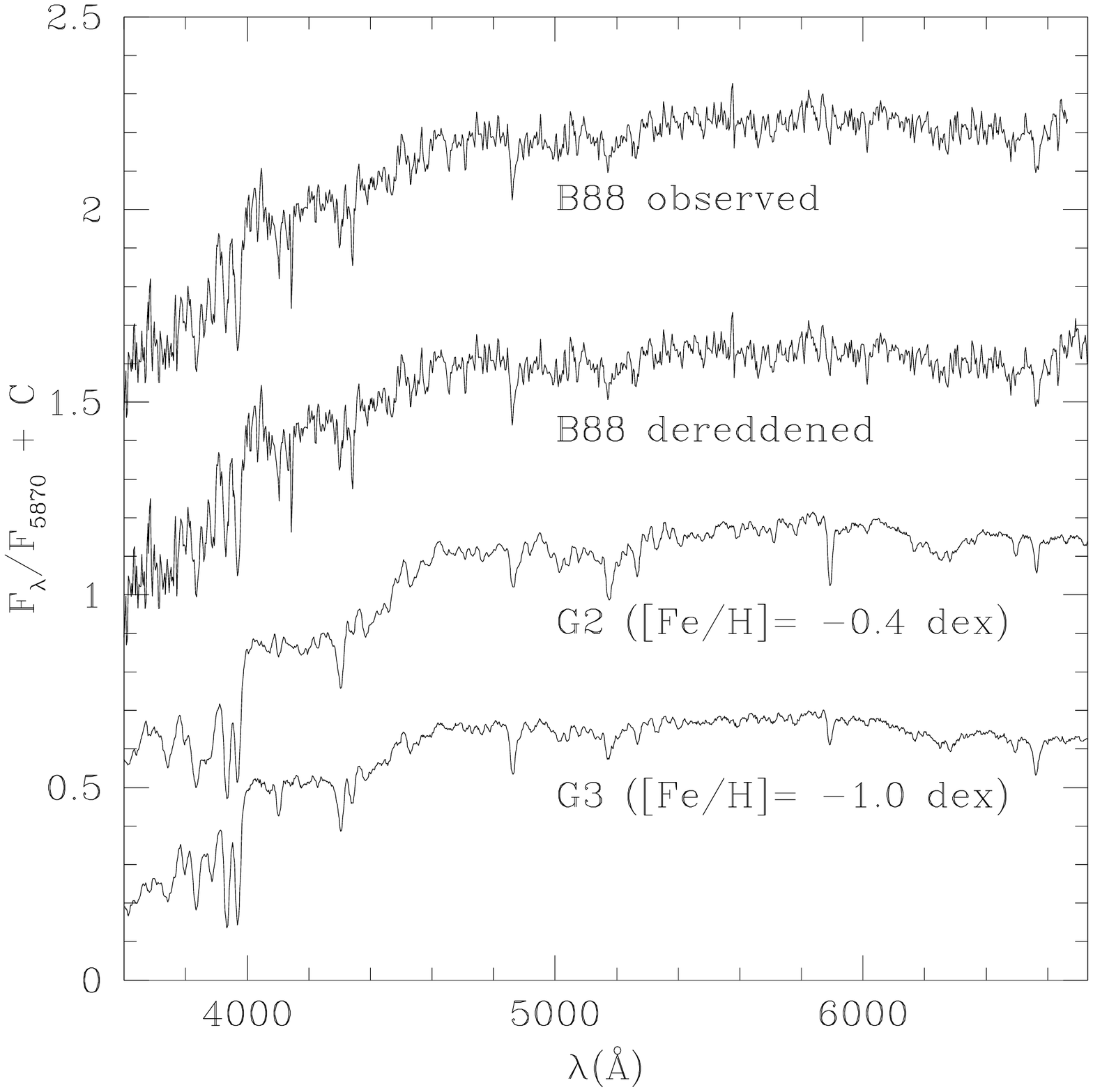}
\caption{Observed and dereddened integrated spectra of Bruck\,88 compared to the G2 and G3
template spectra.}
\label{fig2}
\end{figure}

\begin{figure}
%\centerline{\psfig{figure=fig3.ps,width=84mm}}
\includegraphics[width=84mm]{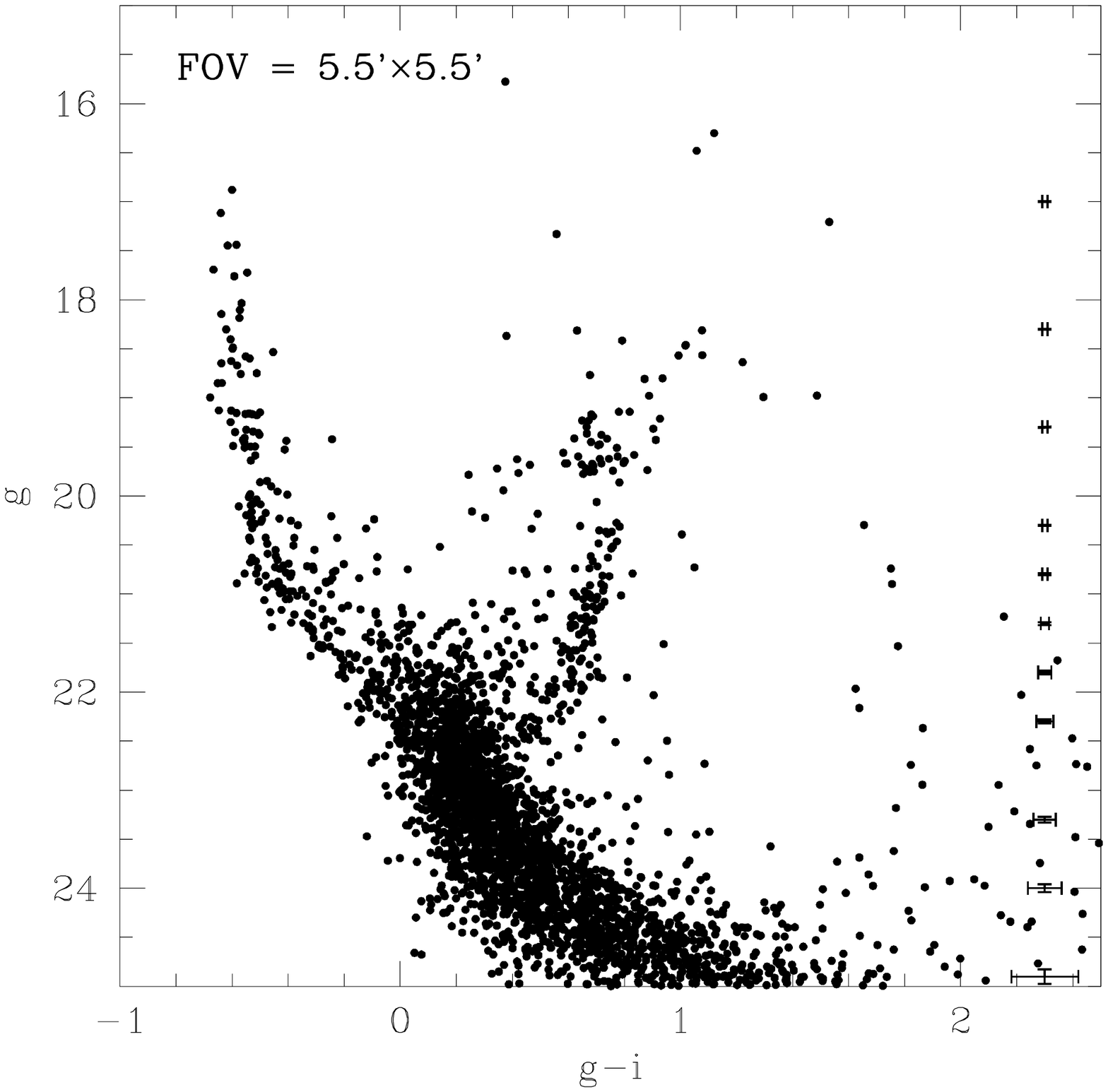}
\caption{Colour-Magnitude diagram of the stars measured in the field of Bruck\,88.
}
\label{fig3}
\end{figure}

\begin{figure}
%\centerline{\psfig{figure=fig4.ps,width=84mm}}
\includegraphics[width=84mm]{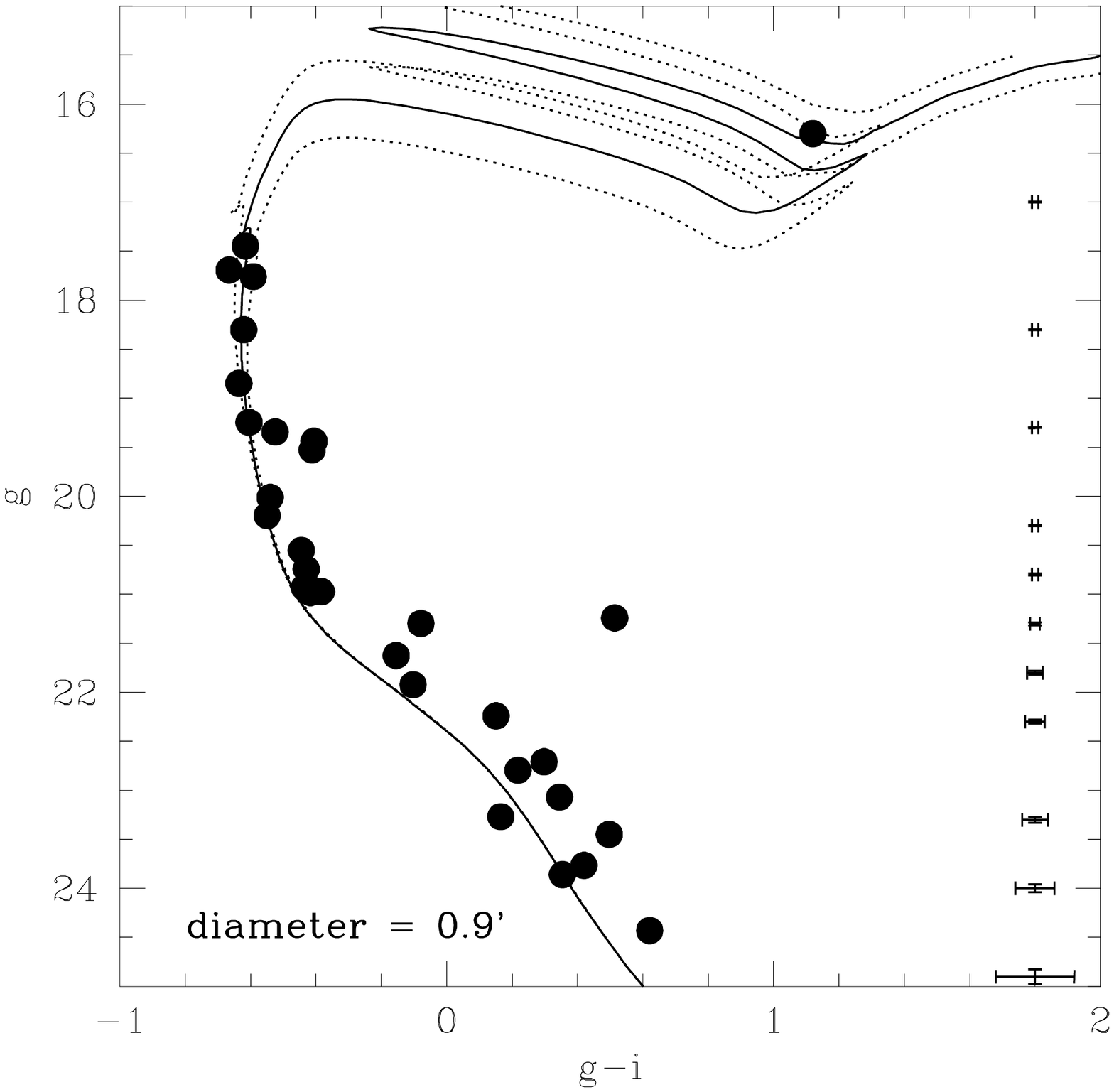}
\caption{Bruck\,88's Colour-Magnitude diagram built using stars distributed within the cluster 
circle.
Theoretical isochrones from Bressan et al. (2012) for log($t$) = 8.0, 8.1, and 8.2, and
metallicity [Fe/H] = -0.7 dex are superimposed.}
\label{fig4}
\end{figure}

\begin{figure}
%\centerline{\psfig{figure=fig5.ps,width=84mm}}
\includegraphics[width=84mm]{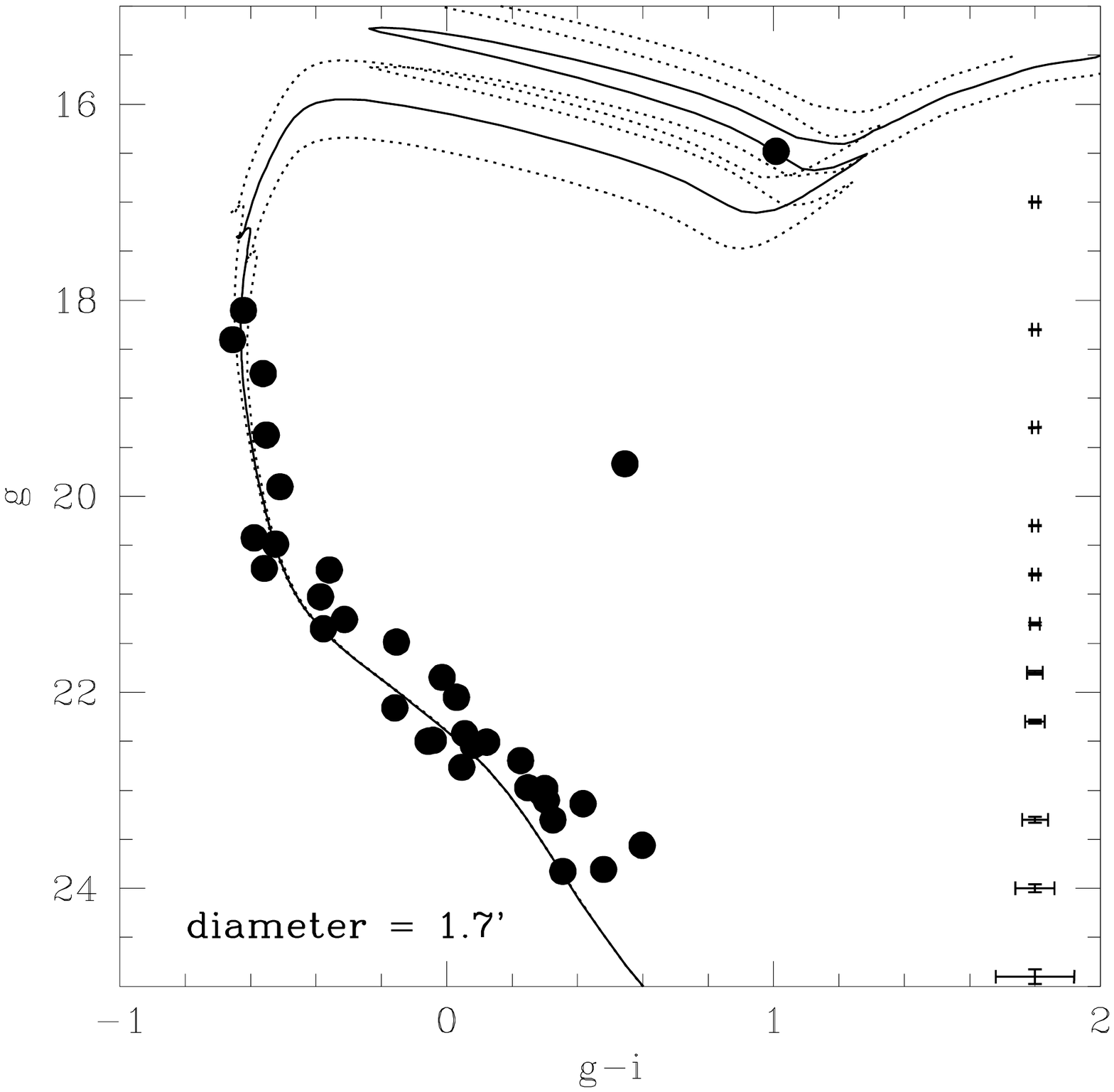}
\caption{Same as Fig. 4 for HW\,33.}
\label{fig5}
\end{figure}

\begin{figure}
%\centerline{\psfig{figure=fig6.ps,width=84mm}}
\includegraphics[width=84mm]{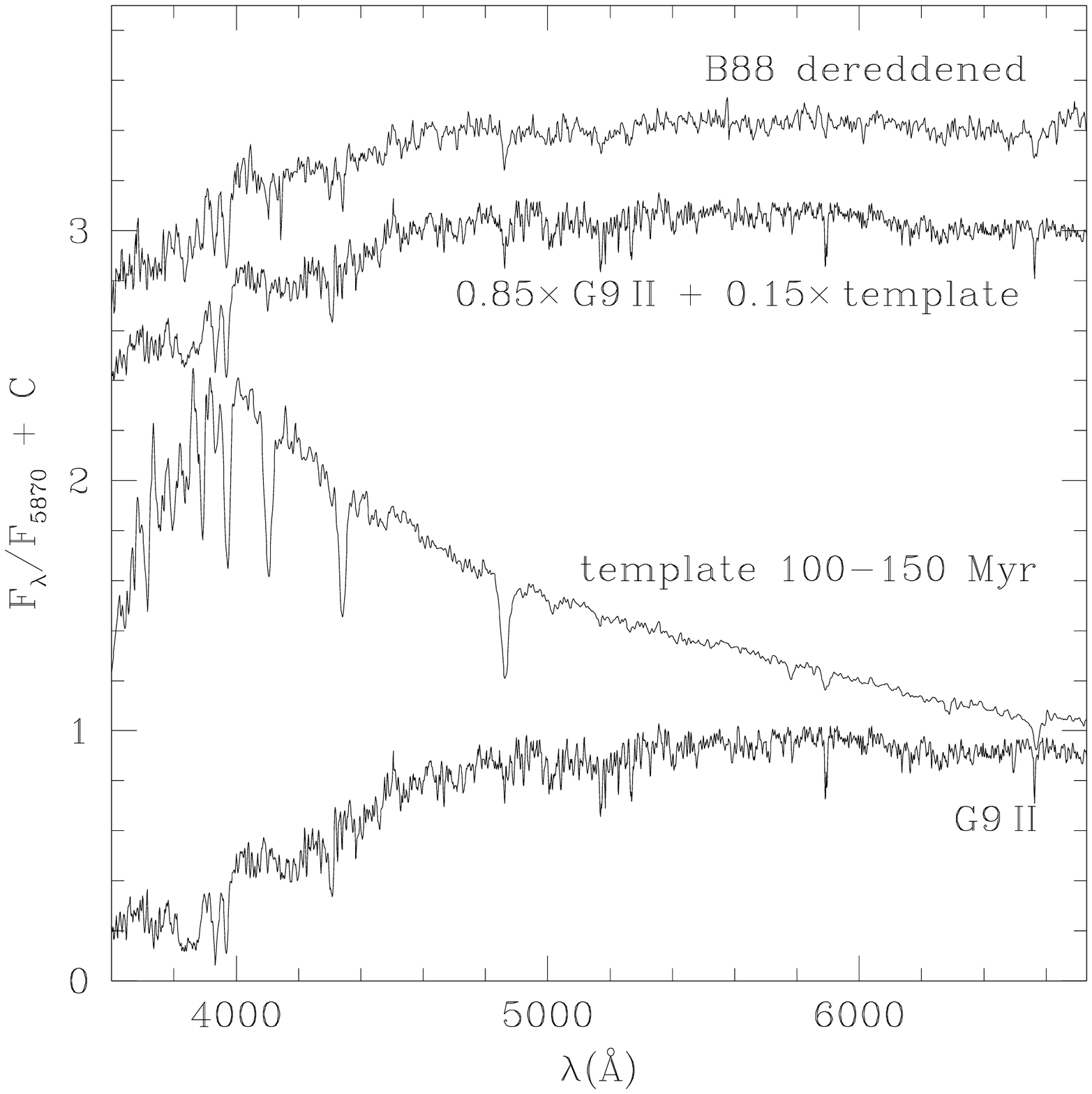}
\caption{Dereddened integrated spectra of Bruck\,88 compared to that resulting
from the combination of the template 100-150 Myr integrated spectrum and of a G9\,II MK type star spectrum.}
\label{fig6}
\end{figure}

\begin{figure}
%\centerline{\psfig{figure=fig7.ps,width=84mm}}
\includegraphics[width=84mm]{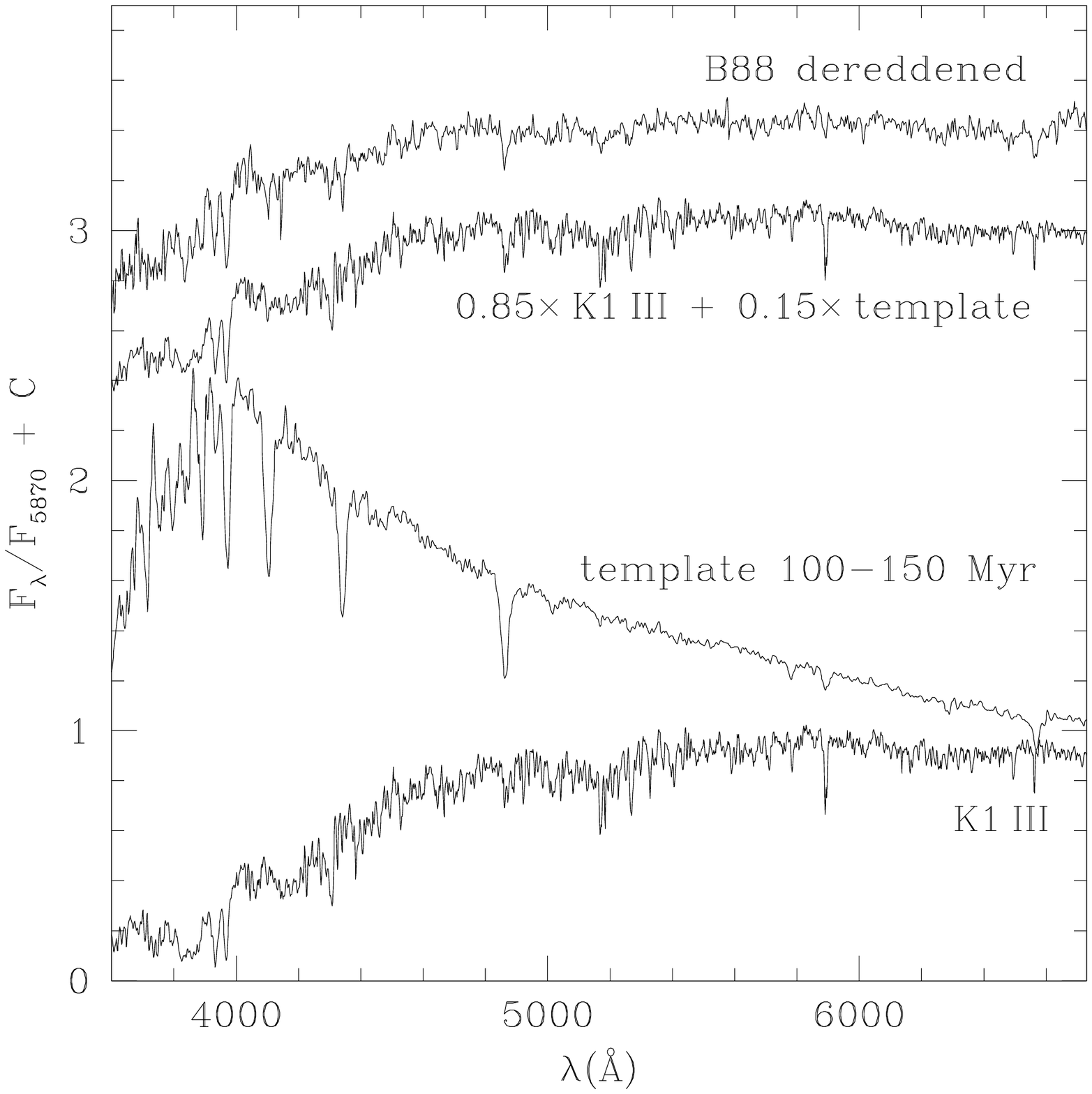}
\caption{Same as Fig. 6 for a K1\,III MK type star spectrum.}
\label{fig7}
\end{figure}

\begin{figure}
%\centerline{\psfig{figure=fig8.ps,width=84mm}}
\includegraphics[width=84mm]{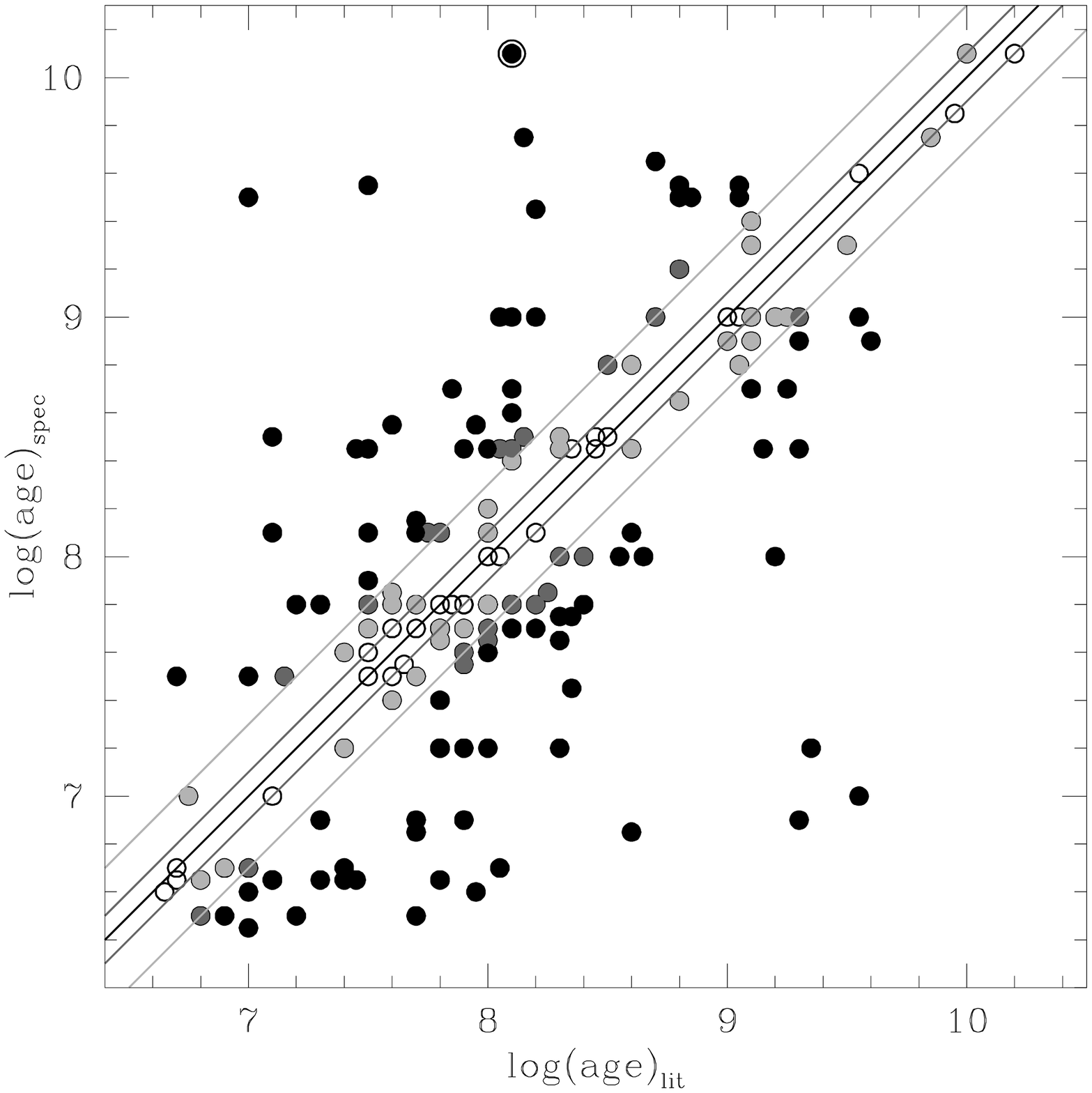}
\caption{Comparison of age estimates derived from the matching of
integrated spectra with those from the literature.}
\label{fig8}
\end{figure}

\begin{figure}
\includegraphics[width=84mm]{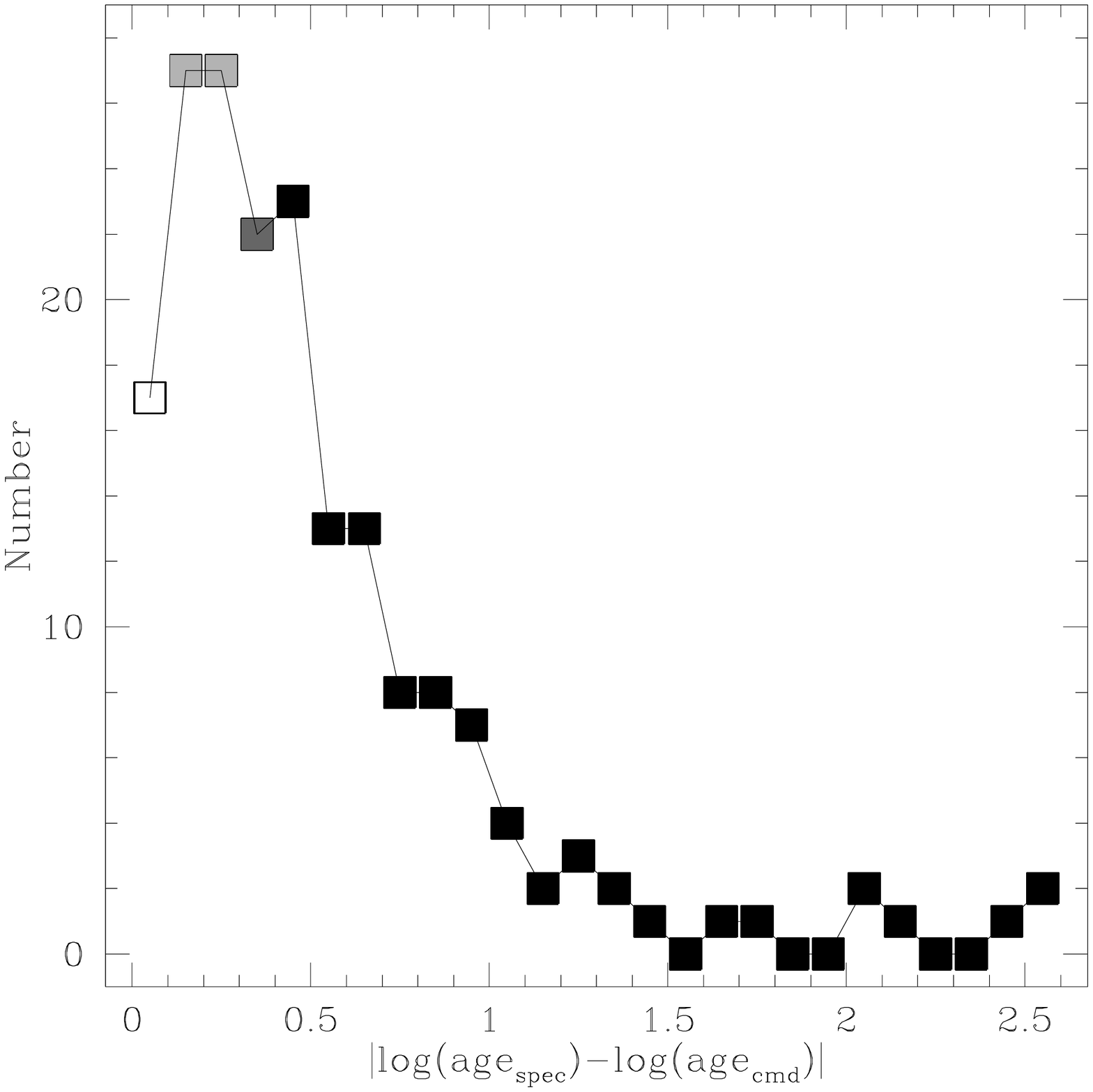}
\caption{Distribution of the difference between ages derived from the
matching of integrated spectra and those from  CMDs (absolute values).}
\label{fig9}
\end{figure}

\begin{figure}
\includegraphics[width=84mm]{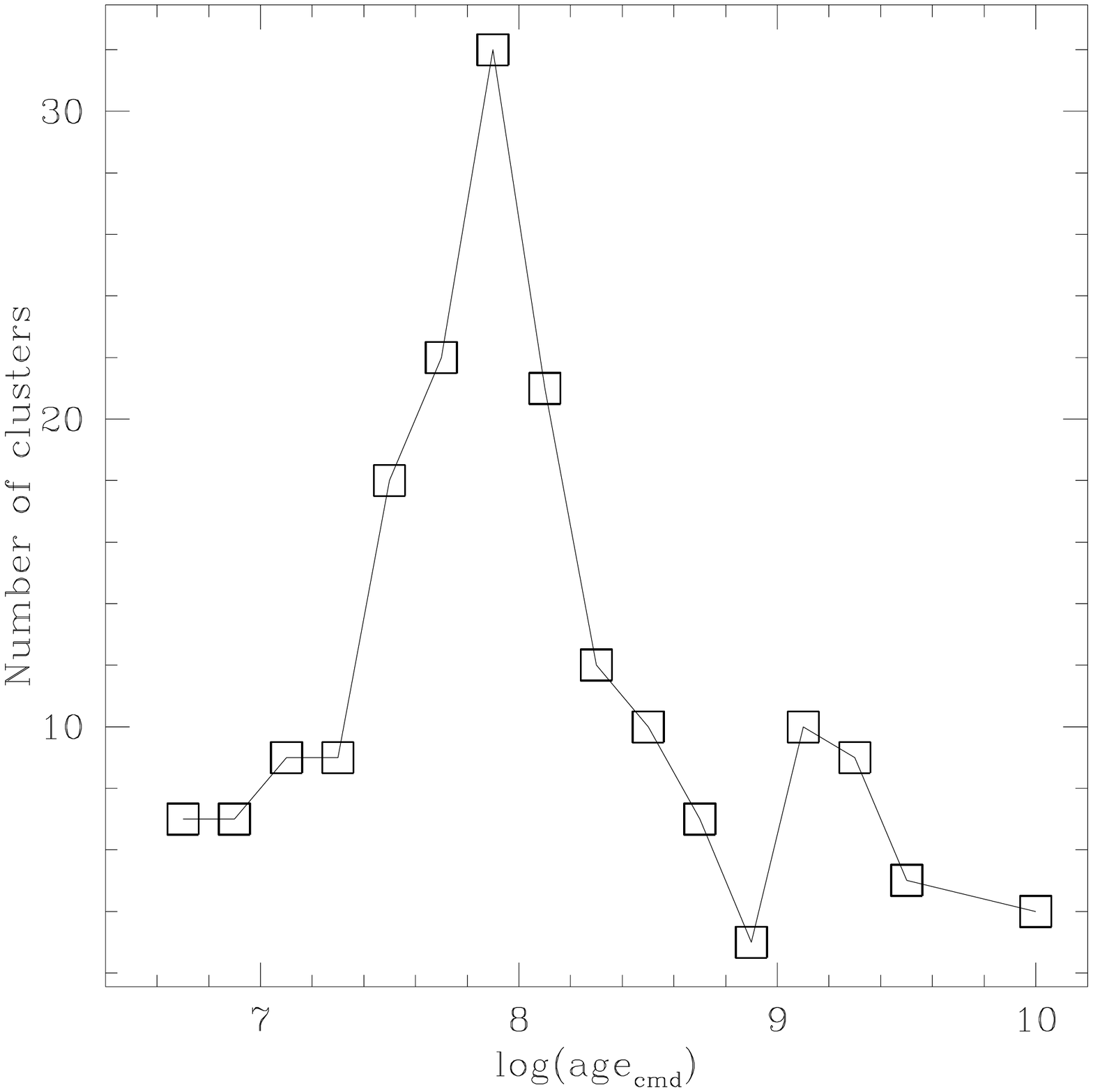}
\caption{Distribution of clusters of Table 3 as a function of 
age (CMD values).}
\label{fig10}
\end{figure}

\begin{figure}
\includegraphics[width=84mm]{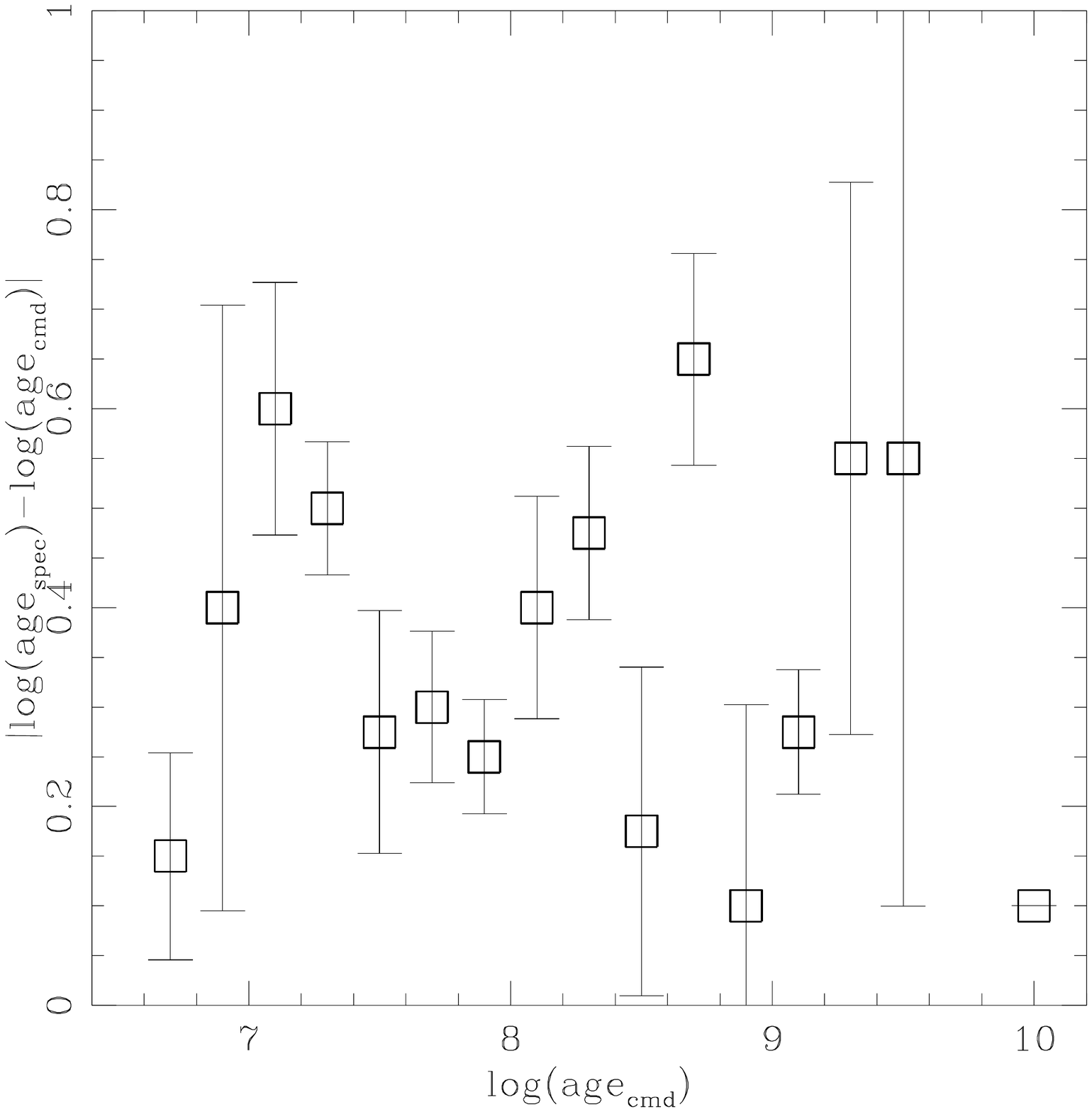}
\caption{Age difference (absolute value) between
integrated spectra and CMD values as a function of cluster
age (CMD value).}
\label{fig11}
\end{figure}

\label{lastpage}
\end{document}